\documentclass{article}
\usepackage{amsmath}
\usepackage{tikz}

\usetikzlibrary{arrows, positioning,arrows.meta}
\usetikzlibrary{backgrounds}
\usetikzlibrary{calc}

\usepackage{svg}
\usepackage{array} 

\usepackage{graphicx}
\usepackage{newtxtext}
\usepackage{newtxmath}
\usepackage[numbers]{natbib} 
\usepackage{hyperref}

\usepackage{authblk} 
\usepackage{enumitem}

\usepackage{bm}
\usepackage[caption=false,justification=raggedleft]{subfig}
\usepackage{algorithm}
\usepackage{algpseudocode}

\usepackage{placeins}

\usepackage[abs]{overpic}  
\makeatletter
\newenvironment{breakablealgorithm}
{
		\begin{center}
			\refstepcounter{algorithm}
			\hrule height.8pt depth0pt \kern2pt
			\renewcommand{\caption}[2][\relax]{
				{\raggedright\textbf{\fname@algorithm~\thealgorithm} ##2\par}%
				\ifx\relax##1\relax 
				\addcontentsline{loa}{algorithm}{\protect\numberline{\thealgorithm}##2}%
				\else 
				\addcontentsline{loa}{algorithm}{\protect\numberline{\thealgorithm}##1}%
				\fi
				\kern2pt\hrule\kern2pt
			}
		}{
		\kern2pt\hrule\relax
	\end{center}
}
\makeatother
\title{\Large Quantized local reduced-order modeling  in time (ql-ROM)}

\author[1,2]{Antonio Colanera}
\author[1,2,3]{Luca Magri\thanks{Email: luca\_magri@polito.it}}
\affil[1]{Department of Mechanical and Aerospace Engineering, Politecnico di Torino, Turin, Italy}
\affil[2]{Department of Aeronautics, Imperial College London, London, United Kingdom}
\affil[3]{The Alan Turing Institute, London, United Kingdom}

\date{}

\begin{document}
\maketitle


\vspace{10pt}

\begin{abstract}
Spatiotemporally chaotic systems, such as the solutions of some nonlinear partial differential equations, are dynamical systems that evolve toward a lower dimensional manifold. 
This manifold has an intricate geometry with heterogeneous density, which makes the design of a single (global) nonlinear reduced-order model (ROM) challenging.  
In this paper, we turn this around. 
Instead of modeling the manifold with one single model, we partition the manifold into clusters within which the dynamics are locally modeled. This results in a quantized local reduced-order model (ql-ROM), which consists of 
(i) quantizing the manifold via unsupervised clustering;  
(ii) constructing intrusive ROMs for each cluster; and 
(iii) seamlessly patch the local models with a change of basis and assignment functions. 
We test the method on two nonlinear partial differential equations, i.e., the Kuramoto-Sivashinsky and 2D Navier-Stokes equations (Kolmogorov flow), across bursting, chaotic, quasiperiodic, and turbulent regimes. 
The local models are built via Galerkin projection onto the local principal directions, which are centered on the cluster centroids. The dynamics are modeled by switching  a local ROM based on the cluster proximity.
The proposed ql-ROM framework has three  advantages over global ROMs (g-ROMs): 
(i) numerical stability, 
(ii) improved short-term prediction accuracy in time, and 
(iii) accurate prediction of long-term statistics, such as energy spectra and probability distributions. The computational overhead is minimal with respect to g-ROMs.
The proposed framework retains the interpretability and simplicity of intrusive projection-based ROMs, whilst overcoming their limitations in modeling complex, high-dimensional, nonlinear dynamics.

\end{abstract}

	\section{Introduction}
The dynamics of high-dimensional dynamical systems, such as those governed by partial differential equations, often evolve chaotically, both in space and time  \cite{Holmes1996, Pope2000}. Chaotic behavior naturally appears in dissipative systems, in which the trajectories tend to evolve to a lower-dimensional manifolds embedded in the high-dimensional state space \cite{strogatz1994book, Fefferman2016}. Accurate modeling of chaotic  dynamics is essential for prediction \cite{noack2003,Rowley2004,NOACK_PAPAS_MONKEWITZ_2005,Racca2023,taira_AIAA2017}, control \cite{BARBAGALLO2009,Brunton2015,Rowley2017,Taira2020}, and real-time data assimilation \cite{Evensen1994,Novoa2022,Novoa2024}, 
 particularly when high-fidelity simulations are computationally expensive \cite{Slotnick2014, Kim1987}. Reduced-order modeling (ROM) is a modeling strategy that reduces the computational cost whilst capturing the behavior of the system with some degree of approximation \cite{Quarteroni2011}.

In recent years, nonlinear ROMs have been developed for  approximating high-dimensional systems with the identification of the underlying solution manifold \cite{Farzamnik2023,Racca2023}. ROMs are particularly attractive for tasks such as real-time prediction, control, and filtering, in which both accuracy and computational efficiency are critical \cite{Papaioannou2022, Buzzicotti2021, Racca2022}. A core goal in the ROM construction is to find a compact representation of the manifold \cite{Benner2015}.

ROM approaches can be broadly categorized as either non-intrusive or intrusive. Non-intrusive ROMs are  data-driven, i.e., they bypass the knowledge of the governing equations and rely  on observables. These methods often leverage machine learning techniques \cite{Brunton2020}, with distinct modeling approaches and strengths. Linear methods such as dynamic mode decomposition (DMD) \cite{ Schmid2022} and resolvent analysis \cite{McKeon2010,Hwang2010,towne2018,Herrmann2021} are well-suited for identifying dominant coherent structures and input-output dynamics in weakly nonlinear regimes. In contrast, regression-based models \cite{Udrescu2020}, such as sparse identification of nonlinear dynamics (SINDy), describe the nonlinear behavior by constructing parsimonious models with sparse regression on a predefined library of functions \cite{Brunton2016, Loiseau2018}. Physics-informed neural networks (PINNs) incorporate governing equations directly into the training process, thereby embedding physical constraints and improving generalization from limited data  \cite{Raissi2019,Papaioannou2022, VonSaldern2022, Buzzicotti2021}. Recurrent neural networks, such as long short-term memory (LSTM), capture the long-term temporal dependencies \cite{Hochreiter1997} and have been used to reduce the  dynamics from latent representations and for minimal dataset design \cite{Bucci2023}, although they may struggle with high-dimensional inputs. Transformers, which use self-attention mechanisms to model  temporal attention mechanisms \cite{Vaswani2017}, have shown promise in sequence modeling tasks, but their application in ROMs remains limited due to large data requirements and computational cost. 
Reservoir computing methods, such as echo state networks (ESNs), provide a computationally cheap framework that is particularly effective for modeling  nonlinear and chaotic systems, requiring minimal training effort to model complex temporal dynamics \cite{Doan2021,Racca2022, Oezalp2023, Racca2023}.
These methods are flexible and require less domain knowledge when the governing equations are unknown or are too difficult to manipulate directly. However, they may face challenges on generalizability, robustness, and physical interpretability, particularly in chaotic or highly nonlinear regimes, in which infinitesimal errors  grow exponentially  \cite{Lorenz1963,brunton2020arfm}.

In contrast, intrusive ROMs are derived from the governing equations of the full-order model (FOM), typically by projecting the equations onto a low-dimensional basis. An example is the POD-Galerkin projection, which decomposes the equations (e.g., Navier-Stokes) onto the principal directions computed with proper orthogonal decomposition (POD) \cite{Rowley2004,Stabile2017,noack2003,deng2020jfm,Sanderse2020}. These models offer  interpretability and approximately fulfill the conservation laws. They are also useful in tasks such as model calibration \cite{Duraisamy2019, Xiao2019}, stability analysis \cite{GIANNETTI2007}, sensitivity analysis \cite{Cinnella2011}, and uncertainty quantification \cite{Xiao2019}. Intrusive methods, however,  can be  numerically unstable or inaccurate if the low-dimensional subspace does not adequately capture the complexity of the system \cite{Carlberg2011}.

Both intrusive and non-intrusive ROMs typically construct a global ROM (g-ROM), i.e. a single model that describes the entire solution manifold. For example, POD-Galerkin models approximate a mean value plus components along principal modes \cite{noack2003}. This assumes that the data distribution is unimodal and well-represented by a linear subspace assumption. This assumption often breaks down in chaotic systems in which the solution manifolds  have intricate and non-Gaussian statistics. Similarly, in autoencoder-based intrusive ROMs \cite{Lee2020a}, the entire dataset is embedded into a latent representation, on which the governing equations are nonlinearly mapped. This requires data and training to resolve the heterogenous  dynamics across the different regions of manifold.
To overcome these limitations, local reduced-order models have been proposed. These models replace a single global representation with a collection of local models, each capturing the behavior in a subset of the solution space \cite{Amsallem2008}. 
 Several methods have been introduced for this purpose, including linear embeddings such as local principal component analysis (PCA) \cite{Kambhatla1997, Roweis2000, Zdybal2023}, and nonlinear approaches like local kernel PCA \cite{Deng2013} and local proper generalized decomposition (PGD) \cite{Badias2017}.
The primary distinction among different local ROMs lies in the definition of ``local", which in turn determines how the dataset is partitioned. 
One class of local approaches defines locality in the physical space, using domain decomposition combined with POD-based basis construction \cite{Corigliano2015, Ferrero2018, Bergmann2018}. These methods have also been extended to promote sparsity in the models \cite{Anderson2022} and to deploy spatially local autoencoding techniques \cite{ConstanteAmores2024}. Another concept is the local temporal clustering of the dynamics. For instance, \cite{Ahmed2020} proposed a temporally localized Galerkin ROM for a two-dimensional turbulent flow, and \cite{Chaturantabut2017}  computed   nonlinear terms with temporally localized bases  using the discrete empirical interpolation method (DEIM). Another class of local approaches clusters snapshots based on the latent phase space. Recently, \cite{Denglocal} introduced a method that combines linear clustering on the solution manifold with linearized dynamics around cluster centroids. For POD-Galerkin models, previous works have developed adaptive local basis methods, in which the projection space is dynamically updated based on the state evolution, with applications to fluid-structure-electrostatics interaction \cite{Amsallem2012}, the inviscid Burgers equation with shock waves \cite{Carlberg2014}, bifurcating flows \cite{Cortes2024, Hess2019}, flame dynamics \cite{Huang2023,Peherstorfer2014}, and cardiac electrophysiology \cite{Pagani2018}. However, these adaptive methods might have numerical stability issues  as their global counterparts.  Finally, manifold learning and local charts construction is an approach to build fully data-driven ROMs over manifold patches \cite{Floryan2022}. 

In this work, we propose quantized local ROMs (ql-ROMs), which is a divide-and-conquer strategy for building reduced-order models on manifolds. The approach quantizes the phase space into a collection of local regions via clustering. Each region is associated with a centroid, which is the reference point for building a local POD-Galerkin model. The  local models are adaptively selected based on the cluster proximity in the solution manifold. The ql-ROMs are tested on nonlinear partial differential equations, whose regimes span from quasi-periodic to turbulent and intermittent.
The paper is organized as follows. Section \ref{sec:methods} introduces the methodology, detailing the phase-space quantization process and the construction of ql-ROMs. Section \ref{sec:datasets} presents the numerical test cases used to validate our method, specifically the Kuramoto-Sivashinsky equation in both bursting and chaotic dynamics, and the 2D Navier Stokes equations (Kolmogorov flow) in quasiperiodic and turbulent regimes. Section \ref{sec:results} showcases the results of the study. The papers ends with conclusions in Section \ref{sec:concl}. Appendices contain numerical details. 

	\section{Quantized local reduced order models (ql-ROMs)}\label{sec:methods}

	We consider a dynamical system  governed by a partial differential equation (PDE)
		\begin{equation}\label{eq:dyn}
			\frac{\partial \bm{u}}{\partial t} + \mathcal{N}(\bm{u},t) = 0,\quad\bm{u}\in\mathbb{R}^{N}, \quad \bm{u}(t=0)=\bm{u}_0
		\end{equation}
		where $\bm{u}$ is the state vector of the system, and $\mathcal{N}$ is a spatially discretized nonlinear differential operator, which contains also the boundary conditions. The state, $\bm{u}$, evolves in the spatial domain $\Omega$ and time $t$.  For example, $\bm{u}$ may include velocity, pressure, and temperature, evaluated on a  grid.

	$M$ time-resolved snapshots
    are sampled at time steps, $\Delta t$, so $t^m = m \Delta t$ represents the time instance of the $m$th snapshot. The corresponding snapshot field is denoted $\bm{u}_m = \bm{u}( t^m)$, where $m = 1, \ldots, M$.

\begin{figure}
\centering
\begin{tikzpicture}[node distance=4.5cm, auto]

    \tikzstyle{block} = [rectangle, draw, fill=white!20, text width=5.5cm, text centered, rounded corners, minimum height=3.6cm]
    \tikzstyle{line} = [draw,  -{Latex[scale=1.15]}]
    
    \node [block,minimum height=0.8cm,  node distance=1.1cm] (block1) {Cartography of the manifold };
    \node [block, below of=block1, node distance=2.25cm] (block2) {\vspace{-12pt} \begin{itemize}[left=0pt,labelsep=0.5em,leftmargin=1em]
			    \item Choice of the low dimensional representation of the manifold (linear vs nonlinear space).
                \item Partition of the manifold in clusters.
			\end{itemize}  };
    \node [block, right of=block2, node distance=6.3cm] (block3) {    \vspace{-15pt}  \begin{itemize}[left=0pt,labelsep=0.5em,leftmargin=1em]
                \item Choice between intrusive or fully data-driven
                  \item Choice between linear or nonlinear models
                 \item modeling the transition among clusters 
            \end{itemize}
};

	\node [block,minimum height=0.8cm, right of=block1, node distance=6.3cm] (block4) {Dynamics modeled on the manifold  };

 \node [block, above of=block1, fill=cyan!20,minimum height=1.2cm, node distance=1.6cm] (blockdata) {Data collection};
 
 \node [block, above of=block4, fill=red!20,minimum height=1.2cm, node distance=1.6cm] (blockpred) {Prediction};
		
    \path [line] (blockdata) -- (block1);
 \path [line] (block1) -- (block4);
 \path [line] (block4) -- (blockpred);

\end{tikzpicture}
\caption{ Quantized local reduced order models (ql-ROMs)
}\label{fig:qlromtik}
\end{figure}
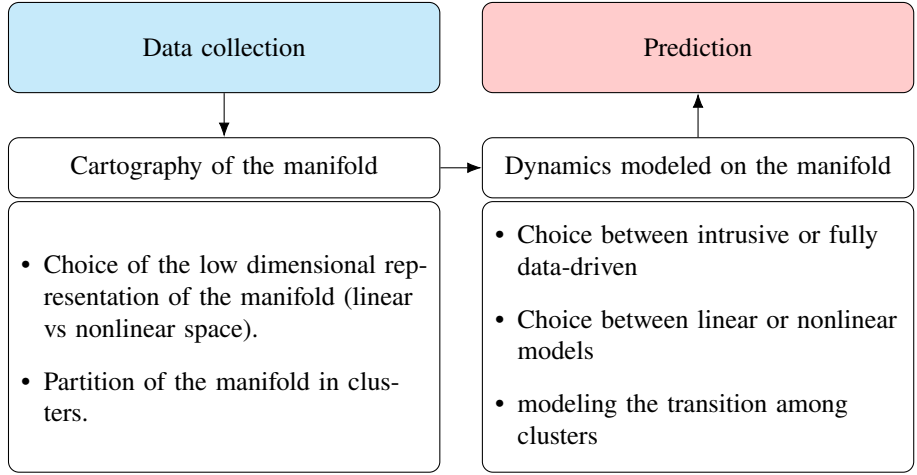

    The objective of reduced order modeling (ROM) is to construct a model with $r \ll N$ degrees of freedom, which accurately captures the essential dynamics of the full order model (FOM)  \cite{Rowley2017}. After a transient, the solution of a dissipative system typically converges to an attractor (solution manifold). In a global reduced-order model (g-ROM), a single ROM is designed to describe the solution manifold. 
	In chaotic systems, however, the attractor has intricate and heterogeneous structures,  making a g-ROM difficult to design, which can lead to numerical instability and large inaccuracy \cite{Floryan2022}. To address this, we propose a quantized local reduced order modeling (ql-ROM) approach in the time domain, which is a divide-and-conquer strategy. We construct $K$ local ROMs, each of dimension $r_k$, which are tailored to different regions of the phase space.
	The proposed method, summarized in Figure \ref{fig:qlromtik}, consists of four stages:  data collection, phase space quantization (section \ref{sec:clu}), the choice of the local ROMs (section \ref{sec:gprom}), and the prediction stage. 

	\subsection{Phase space quantization}\label{sec:clu}

    The first step of ql-ROM consists of creating the cartography of the data manifold by quantizing it into discrete patches (clusters). In this paper, we employ K-means due to its simplicity and computational efficiency. K-means is an unsupervised algorithm that aggregates
similar points (here corresponding to snapshots) into clusters based on a preassigned distance metric.
    Given a dataset of $M$ snapshots, $\bm{u}_m$, in the phase space, the method quantizes this data into $K$ clusters, each centered around a centroid $\bm{c}_k$, where $k = 1, \ldots, K$. These centroids are the barycenters of the clusters and physically represent the mean state of each cluster. The cluster affiliation function, $\beta_v(\bm{u})$, is defined as the function that assign a  point of the phase space $\bm{u}$ to the index of its closest centroid
	\begin{equation}\label{eq
		}
		\beta_v(\bm{u}) = \arg \min_i \|\bm{u} - \bm{c}_i\|, \quad \mathrm{with} \quad i=1,\dots,K,
	\end{equation}
	where $\|\cdot\|$ is a norm. In this work we use the shorthand $\beta(m) := \beta_v(\bm{u}_m)$. Second, we  define a time affiliation function which assigns a time instant $t$ to the cluster of the snapshot that is closest in time

\begin{equation}
    \beta_c(t) = \beta \left( \arg \min_m |t - t_m| \right).
\end{equation}

    The selection of an appropriate distance metric may have an impact on  clustering \cite{Colanera2023,Kelshaw2024}. 	 In this work, we employ the Euclidean metric because of simplicity, i.e., we assume that we have no sufficient prior knowledge on the manifold's shape to justify the choice of different norms.  In the proposed methodology, however, other norms can be chosen without affecting the modeling approach of Figures \ref{fig:qlromtik} and \ref{fig:localROM}. The squared Euclidean distance between two states $\bm{u}_m$ and $\bm{u}_n$  is
	\begin{equation}\label{distFT}
		d_{m,n}^2 = (\bm{u}_m - \bm{u}_n)^{T}(\bm{u}_m - \bm{u}_n),
	\end{equation}
	where $(\cdot)^{T}$ is the transposition operator.
Phase space quantization consists of partitioning the manifold into regions or clusters, each centered around a centroid 
 $\bm{c}_k$, which is defined as the mean of the snapshots within the cluster associated to $\mathbf{c}_k$
	\begin{equation}
	\mathbf{c}_k = \frac{1}{n_k} \sum_{\bm{u}_m \in \mathcal{C}_k} \bm{u}_m = \frac{1}{n_k} \sum_{m=1}^M \chi_k^m \bm{u}_m,
\end{equation}
where $\mathcal{C}_k$ denotes the $k$th cluster and the characteristic function $\chi_i^m$ is
	\begin{equation}
		\chi_i^m =
		\begin{cases}
			1, & \text{if } i = \beta_v(\bm{u}_m). \\
			0, & \text{otherwise}.
		\end{cases}
	\end{equation}
    	The cluster population $n_k$, is the number of snapshots within the $k$th cluster, $n_k = \sum_{m=1}^M \chi_k^m.$
    	Among all possible sets of centroids ${\bm{c}_k}$, we seek for those ${\bm{c}_k^{\star}}$ that 
	\begin{equation}\label{eq
		}
		(\bm{c}_1^{\star}, \ldots, \bm{c}_K^{\star}) = \underset{   \underset{ i=1,\dots,K}{\bm{c}_i} }{\mathrm{arg\,min}} \,J(\bm{c}_1, \ldots, \bm{c}_K),
	\end{equation}
where the objective function, $J$, is the inner-cluster variance
 	\begin{equation}\label{eq:costfun}
		J(\mathbf{c}_1, \ldots, \mathbf{c}_K) = \frac{1}{M} \sum_{m=1}^M \|\bm{u}_m - \mathbf{c}_{\beta(m)}\|^2. 
	\end{equation}
    In the remainder of the paper, we will refer to the optimal centroids $\bm{c}_k^{\star}$ to as $\bm{c}_k$  for brevity.

	To solve the optimization problem \eqref{eq
	}, Lloyd iterations and k-means++ initialization \cite{steinhaus1956,MacQueen1967,Lloyd1982,Arthur2007} are employed. The computational cost of K-means scales almost linearly with the dimension of the state vector $d$, being of the order $O(M K d)$ \cite{Bishop2019,Bloemer2016}.

Although the methodology presented here is compatible with other clustering algorithms, we choose K-means due to its simplicity and efficiency. However, alternative clustering techniques, such as hierarchical clustering \cite{Jain1999}, modularity optimization methods \cite{Newman2006} or density-based approaches like DBSCAN \cite{Ester1996ADA}, could be employed depending on the characteristics of the dataset and application requirements.

	\subsection{Quantized local reduced order models}\label{sec:gprom}
	
Once the phase space has been quantized into $K$ clusters $\mathcal{C}_k$, we design the local reduced order models for each cluster. The approach is a divide-and-conquer approach. First,  we develop a  model, which accurately and locally describes the dynamics within each cluster. Second, we adaptively select the most accurate local model depending on which portion of the attractor the state is.  

\begin{figure}
\centering
\begin{tikzpicture}[node distance=0.5\textwidth, auto]

    \tikzstyle{block} = [rectangle, draw, fill=white!20, text width=0.46\textwidth, text centered, rounded corners, minimum height=6cm]
    \tikzstyle{line} = [draw,  -{Latex[scale=1.15]}]
    
    \node [block] (block1) {\includegraphics[trim= 1cm 1cm 0.5cm 0.7cm,clip,width=1\columnwidth]{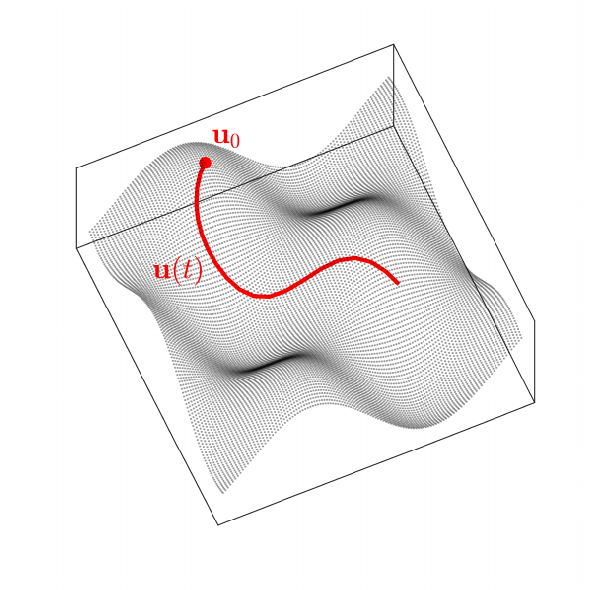}};
    \node [block, right of=block1] (block2) {\includegraphics[trim= 1cm 1cm 0.5cm 0.7cm,clip,width=1\columnwidth]{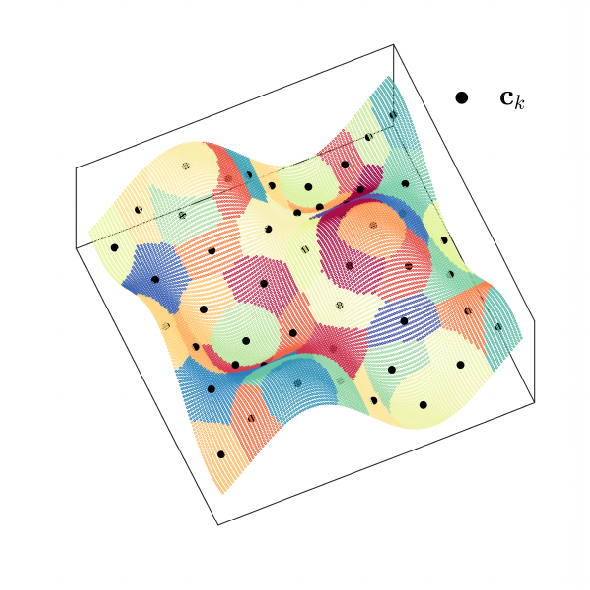}};
    \node [block, below of=block1, node distance=7.2cm] (block3) {\includegraphics[trim= 1cm 1cm 0.5cm 0.7cm,clip,width=1\columnwidth]{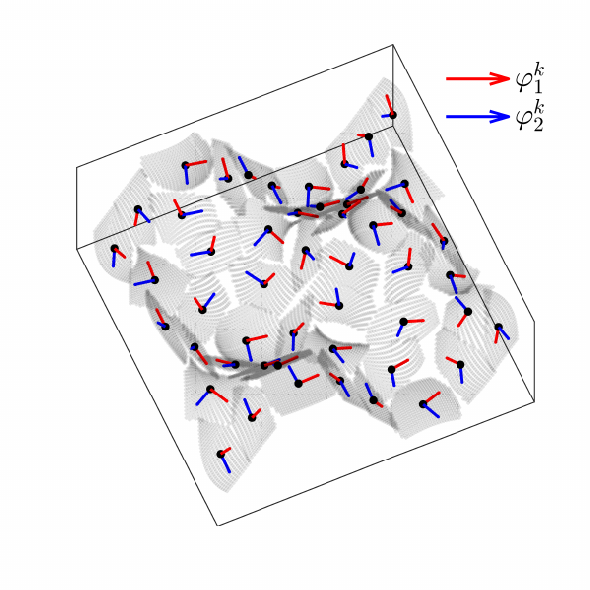}};
    \node [block, right of=block3] (block4) {\includegraphics[trim= 1cm 1cm 0.5cm 0.7cm,clip,width=1\columnwidth]{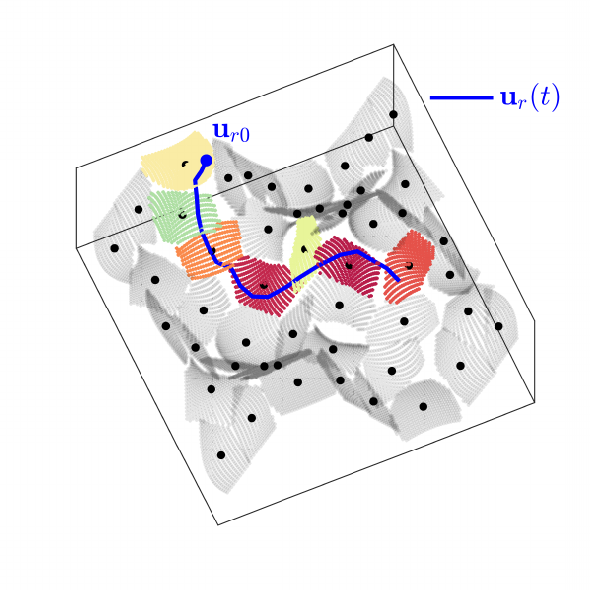}};

 \node [block, above of=block1,minimum height=0.7cm, node distance=3.5cm,text width=0.4\textwidth,xshift=0.03\textwidth] (blockdata) {Data collection};
 \node [block, left of=blockdata,minimum height=0.7cm, node distance=0.245\textwidth,text width=0.03\textwidth] (blockdata) {1};
 \node [block, above of=block2, fill=yellow!20,minimum height=0.7cm, node distance=3.5cm,text width=0.4\textwidth,xshift=0.03\textwidth] (blockdata) {Phase space quantization};
 \node [block, left of=blockdata, fill=yellow!20,minimum height=0.7cm, node distance=0.245\textwidth,text width=0.03\textwidth] (blockdata) {2};
 \node [block, above of=block3, fill=red!20,minimum height=0.7cm, node distance=3.5cm,text width=0.4\textwidth,xshift=0.03\textwidth] (blockdata) {Local basis construction};
 \node [block, left of=blockdata, fill=red!20,minimum height=0.7cm, node distance=0.245\textwidth,text width=0.03\textwidth] (blockdata) {3};
 \node [block, above of=block4, fill=cyan!20,minimum height=0.7cm, node distance=3.5cm,text width=0.4\textwidth,xshift=0.03\textwidth] (blockdata) {Prediction};
\node [block, left of=blockdata, fill=cyan!20,minimum height=0.7cm, node distance=0.245\textwidth,text width=0.03\textwidth] (blockdata) {4};		
\end{tikzpicture}
\caption{
Schematic overview of quantized local reduced order modeling (ql-ROMs). The manifold pictorially represents a high-dimensional attractor on which the solution of the dynamical system lives. The method consists of four stages: (1)~data collection, i.e. trajectories within the state space are collected; (2)~phase space quantization, i.e. the solution manifold is clustered; (3)~local basis construction, i.e. quantized local ROMs are built in cluster centroids (illustrated by local 2D patches); (4)~prediction, i.e. the ROM is deployed to make predictions.
}\label{fig:localROM}
\end{figure}
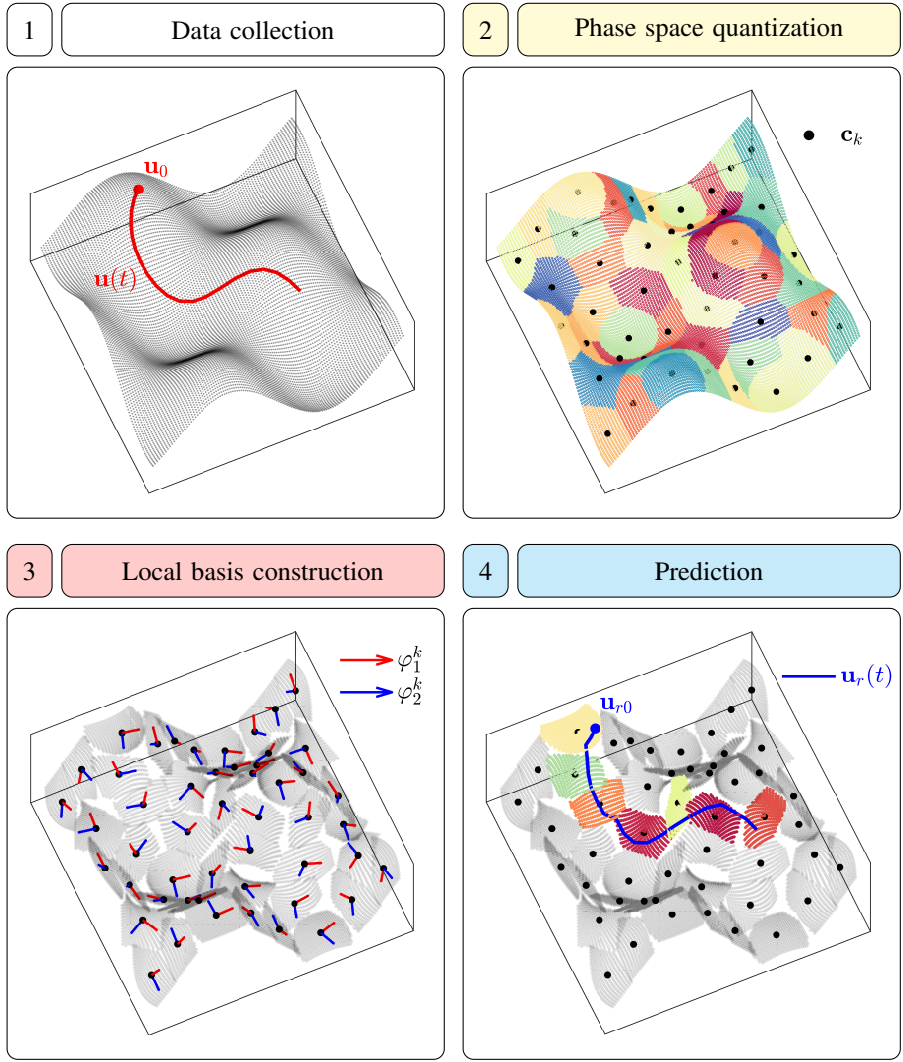

We utilize intrusive deterministic Galerkin proper orthogonal decomposition (Galerkin-POD) ROMs. This approach is simple to implement and offers interpretability. A summary of the proposed ql-ROM methodology is shown in Figure~\ref{fig:localROM}.

\subsubsection{Deterministic local Galerkin ROMs}
For each cluster $k$, we design a local ROM based on the snapshots that belong to that cluster. The local ROM is constructed using the POD snapshot method \cite{Berkooz1993}, which identifies the most energetic modes, in an $L_2$ norm sense, within the cluster. First, we need to compute the fluctuations around the nearest centroid (mean) $\bm{c}_{\beta(m)}$
\begin{equation}\label{eq:fluct}
	\bm{u}'_m = \bm{u}_m -  \bm{c}_{\beta(m)}.
\end{equation}
The centroids are the local means, around which the dynamics of the fluctuations evolves.
Second, the fluctuations snapshots $\{\bm{u'}_m : \beta_v(\bm{u}_m) = k\}$ are used to form the $K$ snapshot matrices
\begin{equation}
	\mathbf{Q}'_k = [\bm{u}'_{m_{1_k}}, \bm{u}'_{m_{2_k}}, \ldots, \bm{u}'_{m_{n_k}}], \quad k=1,\dots,K,
\end{equation}
where $m_{i_k}$ ($i_k=1_k,\dots,n_k$) is the index of the snapshot belonging to cluster $k$. The POD modes are obtained by performing a singular value decomposition (SVD) on $\mathbf{Q}'_k$
\begin{equation}\label{eq:svd}
	\mathbf{Q}'_k = \mathbf{U}_k \bm{\Sigma}_k \mathbf{V}_k^H,
\end{equation}
where $\mathbf{U}_k$ and $\mathbf{V}_k$ contain the spatial and temporal POD modes, respectively, and $\bm{\Sigma}_k$ is a diagonal matrix of singular values. The singular values are arranged in descending order based on their energy content. Third, to construct the quantized local Galerkin ROM,  the state vector $\bm{u}$  at time $t$ is decomposed as
\begin{equation}\label{eq:ansazPOD2}
	\bm{u}( t) = \bm{c}_k + \sum_{i=1}^{r_k} a_i^k(t) \bm{\varphi}_i^k+ \mathrm{truncation\,error}, \quad \mathrm{with} \quad k=\beta_c(t)
\end{equation}
where $\bm{\varphi}_i^k $ is the $ i$th mode, and $a_i^k(t) $ is the temporal coefficient. The number of modes $r_k$ may be different across different clusters, i.e., it is a user choice. In this work, we have consistently used the same number of modes, $r_k=r$, for all the $K$ clusters. In Eq.~\eqref{eq:ansazPOD2}, $ k $ depends on time $t $ through the $\beta_c (t)$, meaning that the POD decomposition adaptively varies from one cluster to another.

Substituting this reduced representation into the original FOM in \eqref{eq:dyn} and performing a Galerkin projection yield a set of $K$ reduced-order models, which are nonlinearly coupled differential equations
\begin{equation}\label{eq:localROM}
	\frac{d \bm{a}^k}{d t} + \mathbf{B}^k \bm{a}^k + \bm{N}^k(\bm{a}^k,\bm{c}_k ) + \bm{f}^k= 0,\quad \bm{a}^k \in	\mathbb{R}^{r_k} \quad \mathrm{with}\quad  k=1,\ldots,K,
\end{equation}
where $\bm{a}^k \in \mathbb{R}^{r_k}$ is the vector of temporal coefficients of the local POD decompositions for cluster $k$, $\mathbf{B}^k \in \mathbb{R}^{r_k\times r_k}$ is a linear operator, and $\bm{N}^k\in\mathbb{R}^{r_k}$ is the nonlinear operator which couples different modes. 
The term $\bm{f}^k \in \mathbb{R}^{r_k}$ contains the projection of the FOM evaluated at the cluster centroids, along with the projection of external forcings (if any).

The dynamical behavior of the system is characterized by a phase space trajectory that evolves toward, and remain confined within, a low-dimensional attractor. As the manifold has been patched, the  state evolves by transitioning from one cluster to another. This transition is based on the nearest centroid, as determined by the cluster-affiliation function. 
If the affiliation function finds a change in the nearest centroid between time steps  $t_m$  and $ t_{m+1} $, i.e., $ \beta(m+1) \neq \beta(m) $, the model transitions from the ql-ROM centered at $ \bm{c}_{\beta(m)} $ to the ql-ROM centered at the nearest centroid $ \bm{c}_{\beta(m+1)} $.
Therefore, a coordinate transformation is required to represent the state at $t_{m+1}$ in the reduced basis of the new cluster. This transformation maps the discrete reduced solution from the previous cluster representation, where $\beta(m) = i$, to the new representation associated with cluster $\beta(m+1) = j$. For the Galerkin-POD local models, the change of coordinates\footnote{Exactly at the boundary of a cluster, because of the change of coordinates, the solution may be non-differentiable. This issue is not important for the goal of this paper, but it can be eliminated with spline-based smoothing in future work \cite{li2020jfm,Colanera2024a}.} is

\begin{equation}\label{eq:swiclu}
\bm{a}^{j} = \bm{U}_{j}^H\bm{U}_{i} \bm{a}^{i}+\bm{U}_{j}^H(\bm{c}_{i}-\bm{c}_{j}),
\end{equation}
where $\bm{U}_i$ and $\bm{U}_{j}$ are the POD mode matrices, computed in \eqref{eq:svd}, of clusters $\beta(m)=i$ and $\beta(m+1)=j$, respectively. All the matrix multiplications in \eqref{eq:swiclu} are computed offline and stored.

The reduced-order model prediction $\bm{u}_r(t)$ is computed by integrating only the ql-ROM corresponding to the current cluster, which is  selected with the cluster-affiliation function. The model initialization requires an initial condition $\bm{u}_{r0}$, which is projected onto the reduced basis of the nearest cluster $\bm{a}_0^{k_0} = \bm{U}_{k_0}^H(\bm{u}_{r0}-\bm{c}_{k_0})$, where $k_0$ is the index of the centroid of the initial condition. Once the time evolution of the reduced coordinates, $\bm{a}^{k}$, and the corresponding cluster-affiliation sequence are stored, the physical state is obtained
\begin{equation}\label{eq:ansazPOD}
	\bm{u}_r( t) = \bm{c}_k + \sum_{i=1}^{r_k} a_i^k(t) \bm{\varphi}_i^k, \quad \mathrm{with} \quad k=\beta_v(\bm{u}_r(t)).
\end{equation}

When $K=1$, the ql-ROM reduces to the traditional global POD-Galerkin model, which has only one centroid (mean field), and the decomposition in Eq.~\eqref{eq:ansazPOD2} is the classical POD decomposition. The procedure is explained in algorithm \ref{alg:pod}. 

	\begin{breakablealgorithm}\label{alg:pod}
		\caption{Procedure for ql-ROMs with POD-Galerkin projections}
		\begin{algorithmic}
        \State \textbf{Offline part:}
			\State \textbf{Collect snapshots:}
			\State Collect $M$ time-resolved snapshots $\{\bm{u}_m\}_{m=1}^M$ from experimental data or high-fidelity simulations.
			
			\State \textbf{Construct cartography of the manifold:}
			\State Use k-means++ algorithm to cluster the snapshots into $K$ clusters.
			\State Determine centroids $\{\bm{c}_k\}_{k=1}^K$ for each cluster.
			
			\State \textbf{For each cluster:}
			\For{$k = 1$ to $K$}
			\State Compute fluctuations $\bm{u'}_m = \bm{u}_m - \bm{c}_k$ for snapshots in cluster $k=\beta(m)$.
			\State Form snapshot matrix $\mathbf{Q'}_k$ for cluster $k$.
			\State Perform SVD on $\mathbf{Q'}_k$ to obtain POD modes $\{\bm{\varphi}_i^k \quad \mathrm{with} \quad i=1,\ldots,r_k\}$.
			\State Construct local Galerkin POD ROM:
			\State $\bm{u}_k := \bm{c}_k + 
            \bm{U}_{k} \bm{a}^{k}(t) $
			\State Derive reduced-order ODEs for temporal coefficients $\bm{a}^k(t)$:
			\State $\frac{d \bm{a}^k}{d t} + \mathbf{B}^k \bm{a}^k + \bm{N}^k(\bm{a}^k,\bm{c}_k) + \bm{f}^k= 0$
			\State Compute transition mapping $\bm{U}_{j}^H\bm{U}_{i}$ and $\bm{U}_{j}^H(\bm{c}_{i}-\bm{c}_{j})$
            \EndFor
			\vspace{5pt}
			\State \textbf{Prediction:}
            \State Given an initial condition $\bm{u}_{r0}$ initialize the ROM of the closest cluster $k_0$
            \State $\bm{a}_0^{k_0} =\bm{U}_{k_0}^H(\bm{u}_{r0}-\bm{c}_{k_0})$
			\While{System state $\bm{u}_r(t)$ prediction}
			\State Evolve ROM using reduced-order ODEs for cluster $i=\beta_c(t)$.
		\State Store $\bm{u}_r(t)$ and reduced representation $\bm{a}^i(t)$
        
			\If{System state prediction $\bm{u}_r(t)$ transitions to cluster $j$}
			\State Transform coordinates from $\bm{a}^i$ to $\bm{a}^{j}$:
			\State 	$\bm{a}^{j} = \bm{U}_{j}^H\bm{U}_i \bm{a}^i+\bm{U}_{j}^H(\bm{c}_i-\bm{c}_{j})$
			\State Switch to ROM for cluster $j$ and continue evolution.
			\EndIf
            \EndWhile
		\end{algorithmic}
	\end{breakablealgorithm}

The ql-ROMs do not increase the degrees of freedom compared to a g-ROM with the same number of modes. The online computational cost remains almost unaffected by the construction of $K$ different local ROMs.  For example, consider a scenario where a POD-Galerkin model retains $r = 10$ modes, meaning that each time step requires solving a system of dimension 10. As an  example, if the phase space is partitioned into $K = 5$ clusters, with each local ROM retaining $r_k = 10$ modes, since at any given time only one of these ROMs is active, the computational cost per time step remains the same as in the global case.  From a computational perspective, the additional cost originates from the online calculation of the distances from the centroids and the change of basis Eq.~\eqref{eq:swiclu}. This change of basis, however, consists of a matrix-vector multiplication and a shift term related to the change of the centroid, both of which are inexpensive compared to the integration of the ROM itself.  Thus, despite the introduction of multiple local ROMs, the online cost remains nearly identical to that of a single g-ROM, whilst improving accuracy, and enabling numerical stability (see later section~\ref{sec:results}).

		\subsection{Choice of $r$ and $K$}\label{sec:choicerK}
The number of clusters, $K$, and the dimensionality of the ROMs, $r_k$, are hyper parameters that are user-defined to accurately capture the dynamics within the low-dimensional representation. The reconstruction error at \textit{m}-th time instance is
\begin{equation}\label{eq:residuo}
	\bm{r}_m = (\bm{I} - \bm{U}_k\bm{U}_k^T)( \bm{u}_m - \bm{c}_{k}), \quad \mathrm{with} \quad k=\beta(m),
\end{equation}
with $\bm{I}\in \mathbb{R}^{N\times N}$ being the identity matrix.
Unless otherwise specified, the number of modes $r$ is selected for the root mean squared error (MSE) \eqref{eq:residuo} of the test dataset, with $K=1$, to be smaller than a threshold, which is shown case by case in section \ref{sec:results}.

When there is no prior knowledge about the manifold's geometry, the number of clusters $K$ is selected based on the Bayesian information criterion (BIC) \cite{pelleg2002,Schwarz1978,priestley1981}. The BIC score is a tool for model selection among a finite set of candidate models \cite{Konishi2008,Wit2012}
\begin{equation}\label{eq:BICbase}
	\text{BIC} =: n_p \log(M) - 2\ell, \quad \mathrm{with}\quad \ell = \log \prod_m P(\bm{u}^m),
\end{equation}
where $\ell$ denotes the log-likelihood of the model, $n_p$ represents the number of parameters in the model, and $P(\cdot)$ is the the probability of a data point. 
In K-means, cluster data are modeled as $ K $ Gaussian distributions, each characterized by its own mean $ \mathbf{c}_k $, whilst sharing a variance $\sigma$. The number of parameters in this case is $ n_p = K \times N $.  The BIC for the K-means is \cite{Colanera2023}
\begin{equation}\label{eq:BICscore}
	\text{BIC} = M \log(J) + K \log(M) - \frac{2}{N} \sum_{k=1}^K n_k \log\left(\frac{n_k}{M}\right),
\end{equation}
where $J$ is the inner-cluster variance \eqref{eq:costfun}. With the BIC score we can select a good model either by minimizing it or by identifying an elbow in its function of $K$, depending on the data. This approach ensures a balanced trade-off between model complexity and goodness of fit.  Specifically, a large $ J $, that is penalized by the first term of \eqref{eq:BICscore}, implies that the points within a cluster are widely spread, indicating the need for additional clusters to better capture data structures. Conversely, a large $K$, penalized in the second termo of \eqref{eq:BICscore}, risks overfitting the data, which leads to inaccurate PDFs.  BIC score offers a trade-off between these two scenarios.  The rightmost term of \eqref{eq:BICscore} is negligible for high dimensional state vectors ($N\gg1$).

	\section{Numerical testcases}\label{sec:datasets}
    We test ql-ROM on three systems that have rich spatiotemporal nonlinear dynamics: the Kuramoto\discretionary{--}{--}{--}Sivashinsky equation and the two-dimensional Navier-Stokes equations (Kolmogorov flow). 

\subsection{Kuramoto–Sivashinsky equation}\label{sec:KSintro}
\begin{figure}
	\centering
	
	\subfloat[\phantom{a}\hfill\phantom{a}]{\centering
		\includegraphics[trim= 0cm 0cm 0cm 0cm,clip,height=3.2cm]{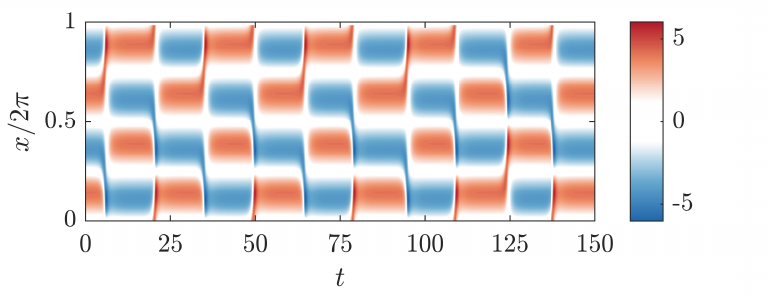}  
		\includegraphics[trim= 0cm 0cm 0cm 0cm,clip,height=3.2cm]{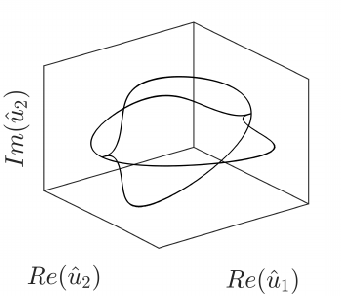}  
	}\\
	\subfloat[\phantom{a}\hfill\phantom{a}]{\centering
		\includegraphics[trim= 0cm 0cm 0cm 0cm,clip,height=3.2cm]{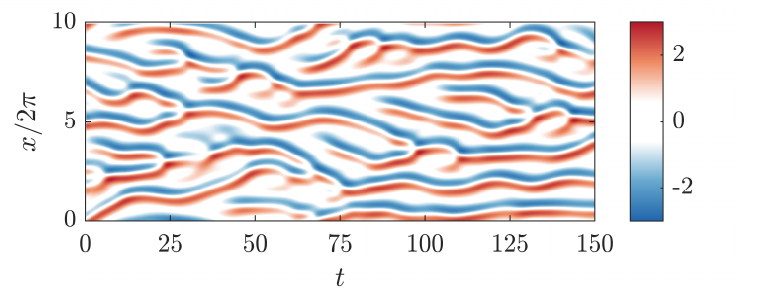}  
		\includegraphics[trim= 0cm 0cm 0cm 0cm,clip,height=3.2cm]{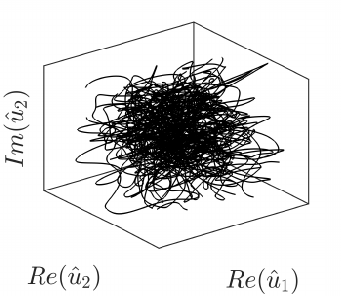}  
	}
	
	\caption{Kuramoto–Sivashinsky equation. Panel (a): Bursting regime for $L=2\pi$ and $\nu=16/71$. Panel (b): Chaotic regime for $L=20\pi$ and $\nu=1$. Left column: Post-transient solution. Right column: Projection of the solution onto the leading spatial Fourier modes $\hat u_i$.}
	\label{fig:ksbase}
\end{figure}

The Kuramoto-Sivashinsky equation (KS) is a nonlinear partial differential equation that describes flame front instabilities \cite{Kuramoto1978,Sivashinsky1977}
\begin{equation}\label{eq:ks}
	\frac{\partial u}{\partial t} + u \frac{\partial u}{\partial x} + \frac{\partial^2 u}{\partial x^2} + \nu \frac{\partial^4 u}{\partial x^4} = 0,
\end{equation}

where $u(x,t)$ is a scalar, $x\in(0, L]$, and the boundary conditions are periodic. Different nonlinear regimes occur as functions of the parameters $L$ and $\nu$ \cite{Hyman1986,Floryan2022}. Figure~\ref{fig:ksbase} shows two regimes of the KS equation. In panel (a), the system has a bursting regime for $L = 2\pi$ and $\nu = 16/71$, characterized by intermittent activity in space and time. In contrast, panel (b) shows a chaotic regime obtained for $L = 20\pi$ and $\nu = 1$, in which the dynamics are chaotic. The left column displays statistically-stationary snapshots of the solution $u(x,t)$, and the right column shows the corresponding projections onto the leading spatial Fourier modes $\hat{u}_i$, providing a compact representation of the behavior in each regime.
Details on the spatial discretization of the numerical solution can be found in Appendix \ref{sec:numerics}. 

\subsection{2D turbulence (Kolmogorov flow)}\label{sec:Kolintro}
The dynamics of fluids are governed by the Navier-Stokes equations, which describe the conservation of mass and momentum of a fluid, respectively:
\begin{align*}
    \nabla \cdot \mathbf{u} &= 0,\\
    	\partial_t \mathbf{u} + (\mathbf{u} \cdot \nabla) \mathbf{u} +\nabla p - \frac{1}{Re} \Delta \mathbf{u} - \mathbf{g} &= 0,
\end{align*}

where $ p $ is the pressure and $ Re $ is the Reynolds number. The velocity field $ \mathbf{u} \in \mathbb{R}^2 $ evolves in the domain $ \Omega = [0, 2\pi)^2$, with periodic boundary conditions enforced on $ \partial \Omega $. A stationary sinusoidal forcing $ \mathbf{g}(\mathbf{x}) = [\sin(4y), 0]^\top$ is imposed on the flow,  where $ y $ is the transverse coordinate \cite{Fylladitakis2018}. This setup, commonly referred to as the Kolmogorov flow \cite{Fylladitakis2018}, generates a nonlinear and multi-scale dataset, which is a benchmark across the turbulent spectrum. To numerically solve this problem, a pseudospectral method has been employed; further details are provided in Appendix~\ref{sec:numerics}.

In this study, two regimes have been analysed, as illustrated in Figure \ref{fig:komolgorovBASE}: 
    \begin{figure}
    \centering
    \subfloat[\phantom{a}\hfill\phantom{a}]{%
        \centering
        \raisebox{-\height}{
                 \begin{overpic}[trim= 0cm 0cm 1.8cm 0cm,clip,width=0.259\columnwidth]{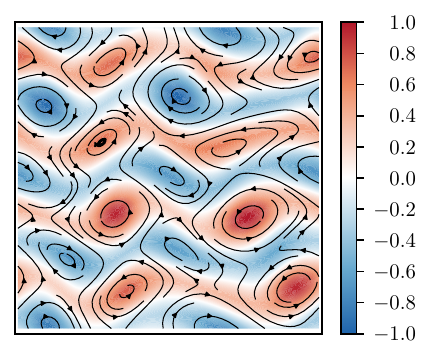}
      \put(-3,50){\small{$y$}} 
       \put(45,-1){\small{$x$}} 
    \end{overpic}

        }
        \quad
        \raisebox{-\height}{\includegraphics[trim= 0cm 0cm 0cm 0cm,clip,width=0.32\columnwidth]{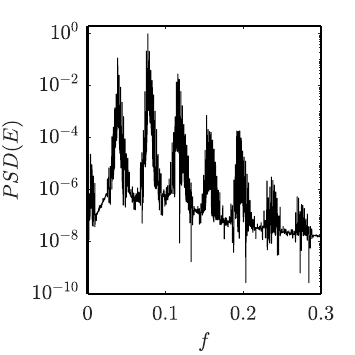}}
      
        \raisebox{-\height}{\includegraphics[trim= 0cm 0cm 0cm 0cm,clip,width=0.32\columnwidth]{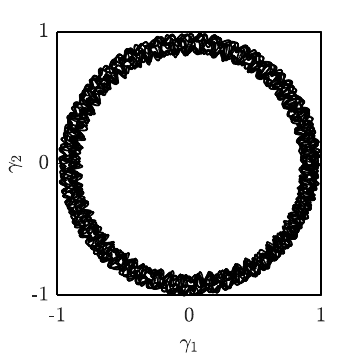}}
    }

    \subfloat[\phantom{a}\hfill\phantom{a}]{%
        \centering
        \raisebox{-\height}{
            \begin{overpic}[trim= 0cm 0cm 1.8cm 0cm,clip,width=0.259\columnwidth]{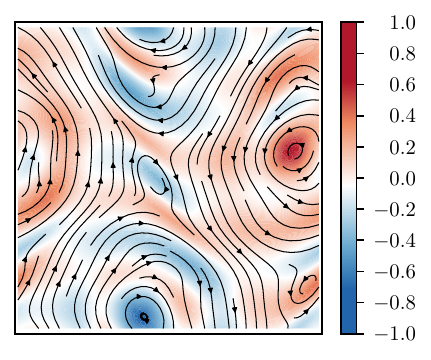}
      \put(-3,50){\small{$y$}} 
       \put(45,-1){\small{$x$}} 
    \end{overpic} 
        }
        \quad
        \raisebox{-\height}{\includegraphics[trim= 0cm 0cm 0cm 0cm,clip,width=0.32\columnwidth]{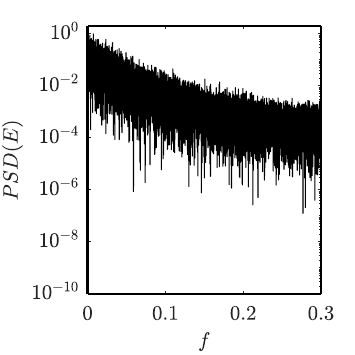}}
        
        \raisebox{-\height}{\includegraphics[trim= 0cm 0cm 0cm 0cm,clip,width=0.32\columnwidth]{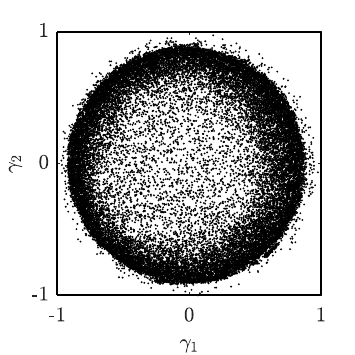}}
    }

    \caption{Kolmogorov flow. Panel (a): $Re = 20$, Panel (b): $Re = 42$. 
        First column: overlay of vorticity ($\nabla \times \bm{u}$) and streamlines. $0 < x < 2\pi$ and $0 < y < 2\pi$. 
        Second column: normalized power spectral density (PSD) of the total kinetic energy. 
        Third column: leading multidimensional scaling (MDS) coordinates for regime visualization purposes.}
    \label{fig:komolgorovBASE}
\end{figure}
a quasiperiodic regime ($ Re = 20 $, panel (a)); and a chaotic and turbulent regime ($ Re = 42 $, panel (b)) \cite{Platt1991}.
The quasiperiodic configuration ($ Re = 20 $) has a cellular pattern, as highlighted also in \cite{Platt1991}, whereas for $ Re = 42 $, the flow becomes both spatially and temporally chaotic.  For $ Re = 20 $, the spectrum is tonal and quasiperiodic, while at $ Re = 42 $, the spectrum becomes broadband, with no dominant frequencies. In the right column, the leading multidimensional scaling (MDS) variables, $ (\gamma_1, \gamma_2) $, are shown. MDS, as detailed in Appendix \ref{sec:MDS}, is a dimensionality reduction technique that preserves pairwise distances between points, making it particularly effective for visualizing intricate flow regimes \cite{kaiser2014}. The coordinates $ (\gamma_1, \gamma_2) $ show the two different topologies of the two regimes.

\section{Results}\label{sec:results}
The ql-ROM is first demonstrated in Section~\ref{sec:resKS} on the Kuramoto–Sivashinsky (KS) equation in both bursting and chaotic regimes. In Section~\ref{sec:resKOL}, the method is demonstrated on the 
two-dimensional Navier-Stokes equations (Kolmogorov flow).
\subsection{Kuramoto–Sivashinsky equation}\label{sec:resKS}
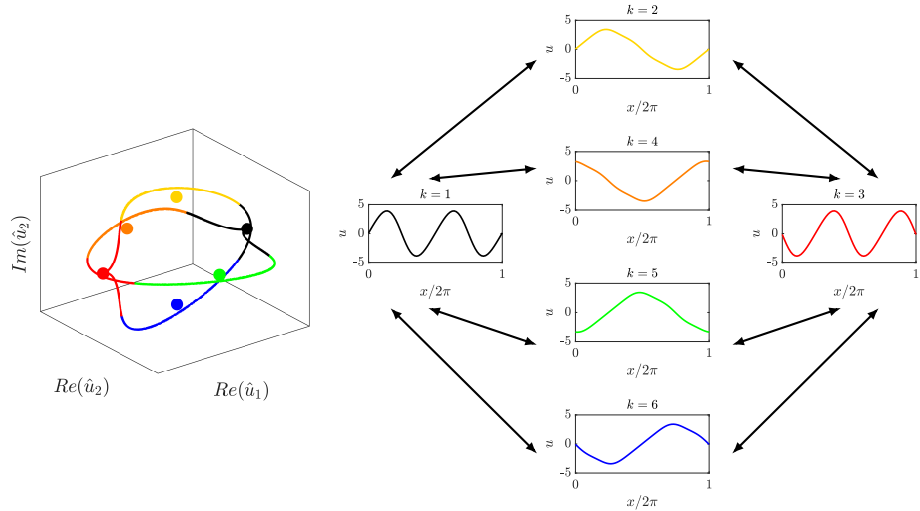
\begin{figure}
	\centering
	\begin{minipage}{0.36\columnwidth}
			\includegraphics[trim= 0cm 0cm 0cm 0cm,clip,height=4cm]{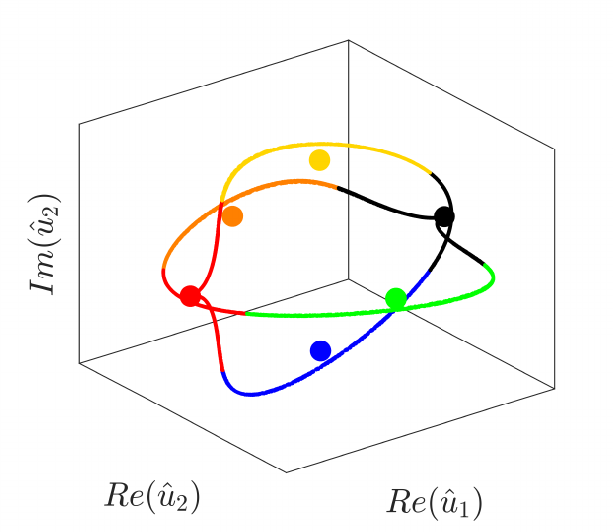}
	\end{minipage}
	\begin{minipage}{0.6\columnwidth}\centering
	\begin{tikzpicture}[node distance=1.75cm, auto]
		
		\tikzstyle{block} = [rectangle, draw=white, rounded corners,  fill=none, text width=2.5cm, rounded corners, minimum height=1.7cm, inner sep=1pt]
		
		\node [block] (block1) { \includegraphics[trim=0cm 0.0cm 0cm 0.cm,clip,height=1.55cm]{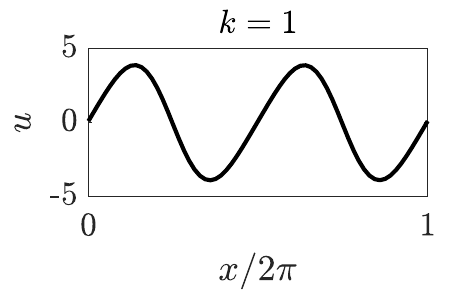}};
		
		\node [block, right of=block1, xshift=1cm,yshift=2.45cm] (block2) {\includegraphics[trim=0cm 0.0cm 0cm 0.cm,clip,height=1.55cm]{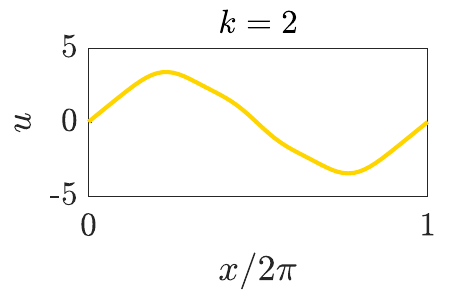}};
		\node [block, below of=block2] (block3) {\includegraphics[trim=0cm 0.0cm 0cm 0.cm,clip,height=1.55cm]{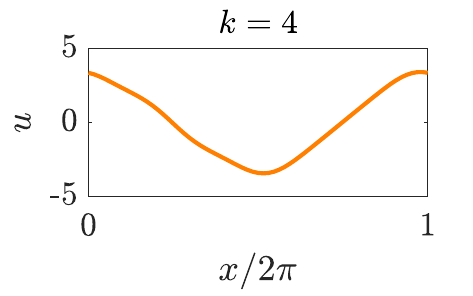}};
		\node [block, below of=block3] (block4) {\includegraphics[trim=0cm 0.0cm 0cm 0.cm,clip,height=1.55cm]{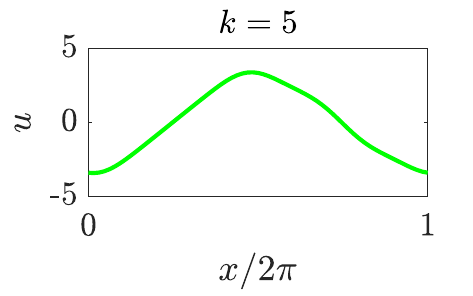}};
		\node [block, below of=block4] (block5) {\includegraphics[trim=0cm 0.0cm 0cm 0.cm,clip,height=1.55cm]{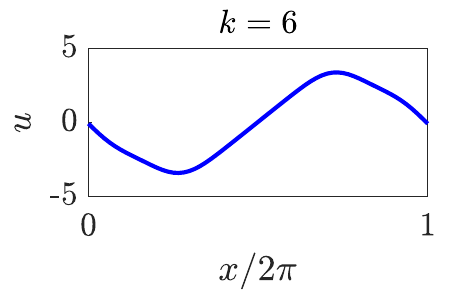}};
		
		\node [block, right of=block2, xshift=1cm, yshift=-2.45cm] (block6) {\includegraphics[trim=0cm 0.0cm 0cm 0.cm,clip,height=1.55cm]{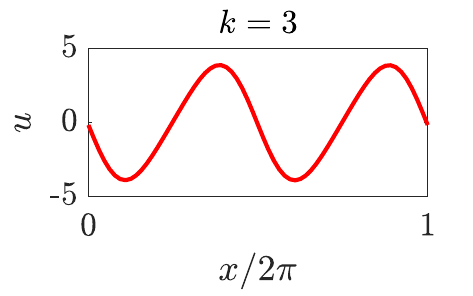}};

		\draw[{Latex[scale=0.7]}-{Latex[scale=0.7]}, thick] ([xshift=-0.5cm]block1.north) -- (block2.west);
		\draw[{Latex[scale=0.7]}-{Latex[scale=0.7]}, thick] (block1.north) -- ([yshift=0.3cm]block3.west);
		\draw[{Latex[scale=0.7]}-{Latex[scale=0.7]}, thick] (block1.south) -- ([yshift=-0.3cm]block4.west);
		\draw[{Latex[scale=0.7]}-{Latex[scale=0.7]}, thick] ([xshift=-0.5cm]block1.south) -- (block5.west);
		
		\draw[{Latex[scale=0.7]}-{Latex[scale=0.7]}, thick] ([xshift=0.5cm]block6.north) -- (block2.east);
		\draw[{Latex[scale=0.7]}-{Latex[scale=0.7]}, thick] (block6.north) -- ([yshift=0.3cm]block3.east);
		\draw[{Latex[scale=0.7]}-{Latex[scale=0.7]}, thick] (block6.south) -- ([yshift=-0.3cm]block4.east);
		\draw[{Latex[scale=0.7]}-{Latex[scale=0.7]}, thick] ([xshift=0.5cm]block6.south) -- (block5.east);
		
	\end{tikzpicture}
		\end{minipage}
	\caption{Kuramoto–Sivashinsky equation in bursting regime. Left panel: Clustered phase space of the  KS equation in the bursting regime. Both the cluster centroids and snapshots are color-coded based on the corresponding cluster affiliations. Right panel: Spatial distribution of clusters centroids.}
	\label{fig:KSbursting}
\end{figure}

The KS equation, \eqref{eq:ks}, has different nonlinear dynamics for different values of the parameters $L$ and $\nu$. In bursting regime, the solution evolves intermittently between pseudo-steady cellular states of opposite sign \cite{Floryan2022}. When projected onto the leading spatial Fourier modes, the system oscillates between two saddle points connected by four heteroclinic orbits. The solution has six distinct regions in the geometry of the  manifold, suggesting a natural choice of $K=6$ clusters, as shown in the left panel of Figure~\ref{fig:KSbursting}. 
The solid line is the trajectory and the markers are the cluster centroids, both color coded according to the cluster affiliation function. The right panel shows the spatial distribution of the six centroids. $k=1$ (black) and $k=3$ (red) centroids represent the metastable states with the remaining centroids being transitional, representing the four heteroclinic orbits. The flow evolves alternatively from 
the one metastable cluster another via transitional clusters.

\begin{figure}
	\centering
\subfloat[\phantom{a}\hfill\phantom{a}]{\centering
\includegraphics[trim= 0cm 0cm 0cm 0cm,clip,width=0.5\columnwidth]{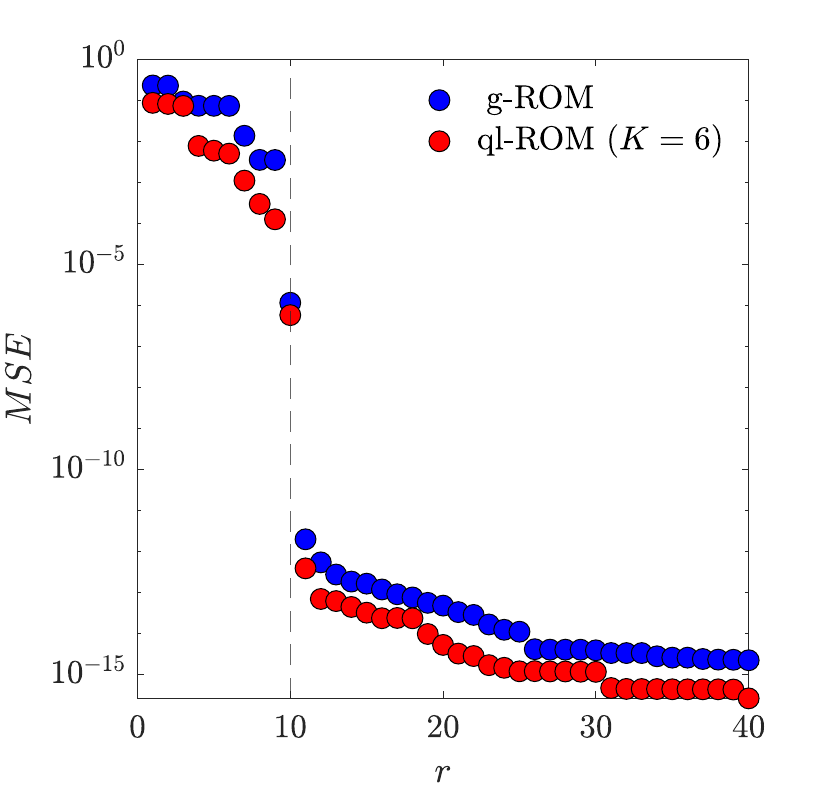}    }
\subfloat[\phantom{a}\hfill\phantom{a}]{
\begin{minipage}{0.45\columnwidth}
\centering
\includegraphics[trim= 0cm 0.cm 0cm 0.cm,clip,width=0.49\columnwidth]{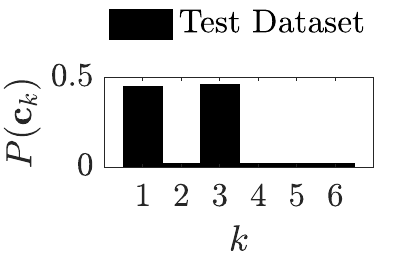}  \\
\includegraphics[trim= 0cm 0.cm 0cm 0.cm,clip,width=0.49\columnwidth]{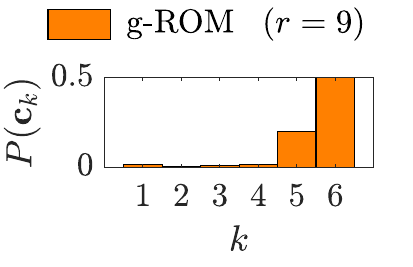} 
\includegraphics[trim= 0cm 0.cm 0cm 0.cm,clip,width=0.49\columnwidth]{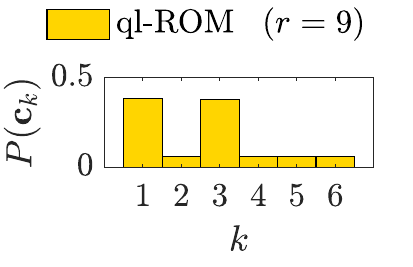}  \\
\includegraphics[trim= 0cm 0.cm 0cm 0.cm,clip,width=0.49\columnwidth]{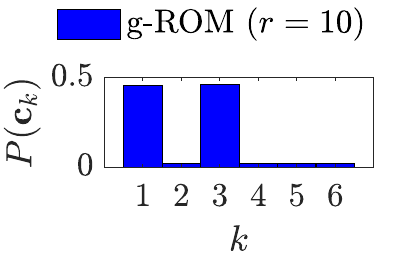} 
\includegraphics[trim= 0cm 0.cm 0cm 0.cm,clip,width=0.49\columnwidth]{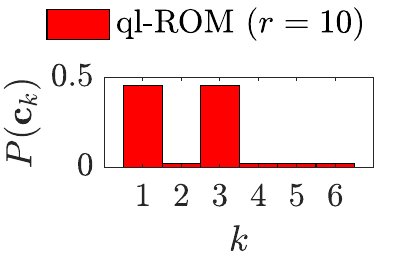}  \\
\end{minipage}
}\\
\subfloat[\phantom{a}\hfill\phantom{a}]{\centering
\includegraphics[trim= 0cm 0cm 0cm 0cm,clip,width=0.5\columnwidth]{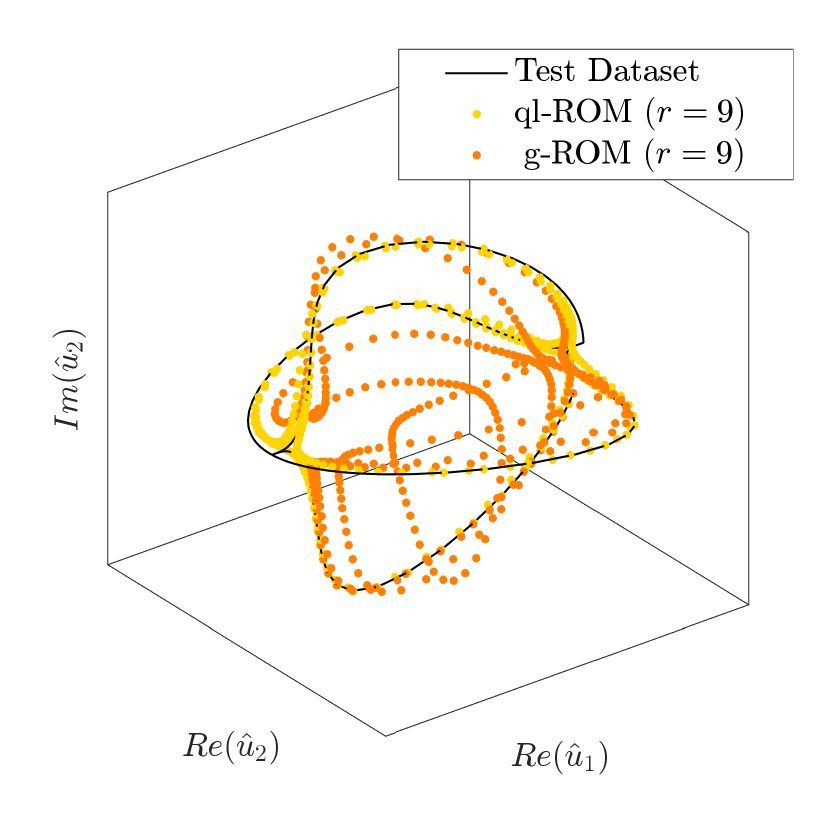}    }    
\subfloat[\phantom{a}\hfill\phantom{a}]{\centering
\includegraphics[trim= 0cm 0cm 0cm 0cm,clip,width=0.5\columnwidth]{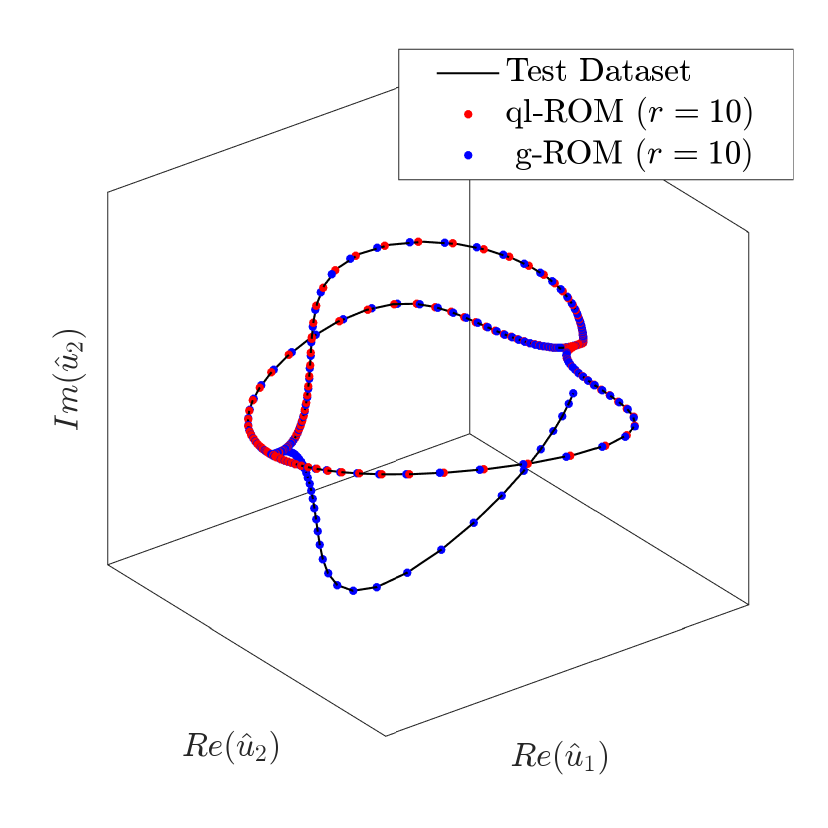}    }

	\caption{Kuramoto–Sivashinsky equation in bursting regime. Panel (a): Reconstruction error of the predicted snapshots using a single POD basis (in blue) and local POD modes with 6 clusters (in red). Both the basis and the centroids were constructed using only the training dataset. Panel (b): Probability distribution of cluster affiliations for the baseline (black), g-ROM ($r=9$) (orange),  ql-ROM ($r=9$, $K=6$) (yellow),  g-ROM ($r=10$) and ql-ROM ($r=10$, $K=6$) (red). Panels (c-d): Phase portrait comparison with $r=9$ and $r=10$. Baseline (black), g-ROM with $r=9$ (orange), ql-ROM with $r=9$ (yellow), g-ROM with $r=10$ (blue) and ql-ROM with $r=10$ (red). With 9 modes, the g-ROM is even unstable, whereas the ql-ROM accurately captures both the $\beta$ statistics and the underlying manifold geometry. With $r = 10$, both approaches successfully capture the geometry and the probability density function of $\beta$.
 }
	\label{fig:ksr9r10}
\end{figure}

Once the solution phase space is clustered in different regions, the local ROMs can be constructed. Panel (a) of Figure~\ref{fig:ksr9r10} shows the MSE of \(\mathbf{r}_m\) \eqref{eq:residuo} on the test dataset for varying \(r\), with a sharp decrease at \(r=10\), which motivates the choice of \(r=10\) as the number of modes for this case.
In Panels (c-d) of Figure \ref{fig:ksr9r10}, show the dynamics prediction of the test dataset for $r=9$ (panel (c)) and $r=10$ (panel (d)). Remarkably, with $r=9$ the g-ROM is  unstable whilst the ql-ROM is stable and accurate.

The probability $P(\bm{c}_k)$ to be in a cluster $k$ can be estimated with
\begin{equation}
	P(\bm{c}_k) = \frac{ n_k}{M}.
\end{equation}
The vector containing all the $P(\bm{c}_k)$ indicates whether the predicted trajectories  populate the phase space similarly to the original data \cite{Hou2022}. Figure \ref{fig:ksr9r10}, panel (b), shows the  probabilities of the clusters affiliation function of the test dataset (black), the g-ROM and the ql-ROM with $r=9$ and $r=10$. With $9$ modes the g-ROM performs poorly, while with $r=10$ both models are accurate.
 
\begin{figure}
	\centering
	\subfloat[\phantom{a}\hfill\phantom{a}]{\centering\includegraphics[trim= 0cm 1cm 0cm 1.5cm,clip,height=10.3cm]{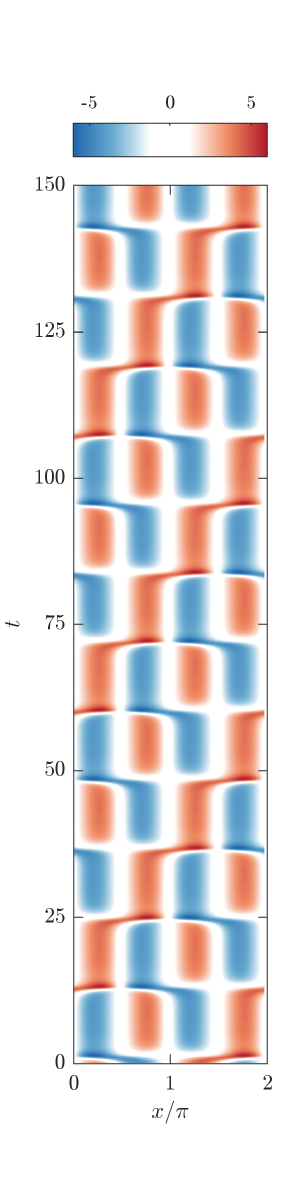}}
	\subfloat[\phantom{a}\hfill\phantom{a}]{\centering\includegraphics[trim= 1.1cm 1cm 0cm 1.5cm,clip,height=10.3cm]{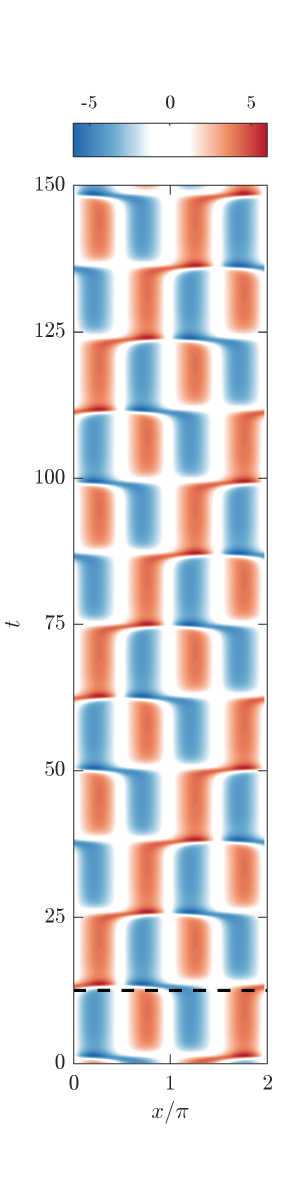}}
	\subfloat[\phantom{a}\hfill\phantom{a}]{\centering\includegraphics[trim= 1.1cm 1cm 0cm 1.5cm,clip,height=10.3cm]{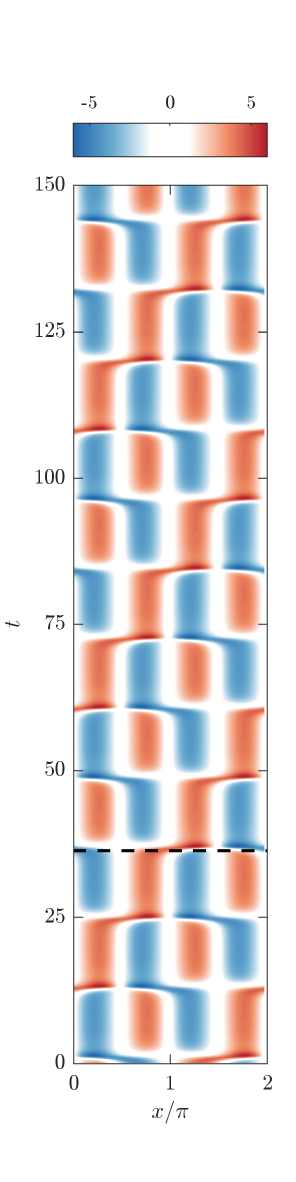}}
	\subfloat[\phantom{a}\hfill\phantom{a}]{\centering\includegraphics[trim= 1.1cm 1cm 0cm 1.5cm,clip,height=10.3cm]{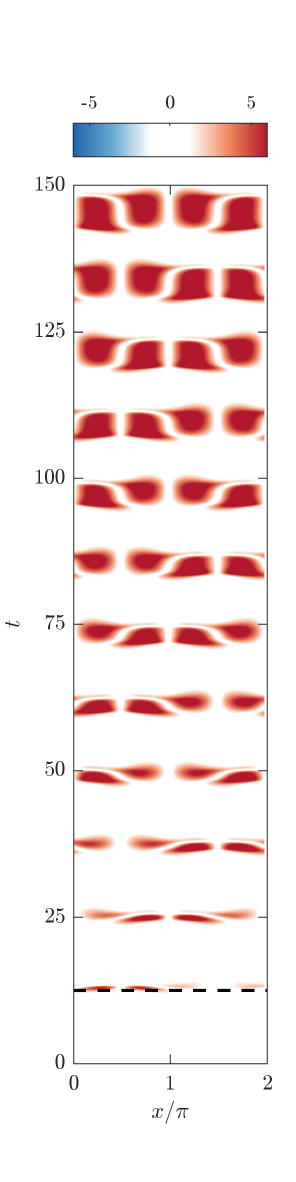}}
	\subfloat[\phantom{a}\hfill\phantom{a}]{\centering\includegraphics[trim= 1.1cm 1cm 0cm 1.5cm,clip,height=10.3cm]{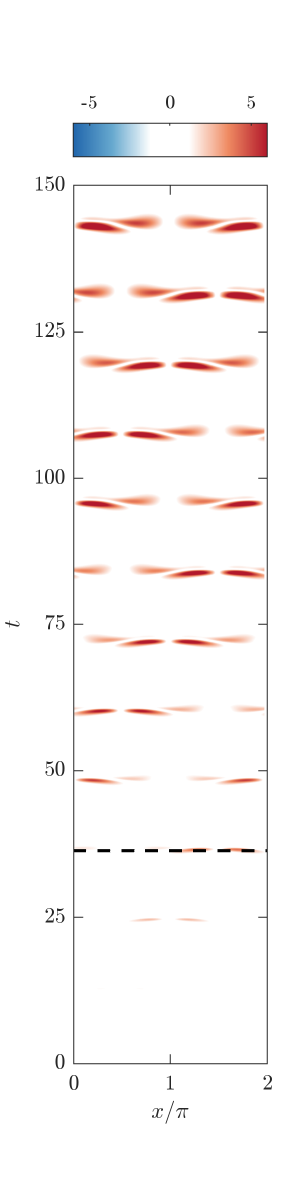}}
	
	\caption{Kuramoto–Sivashinsky equation in the bursting regime. Comparison between the ground truth and the predictions obtained from g-POD and ql-ROMs. Panels (a–c) show the ground truth, the g-ROM, and the ql-ROM predictions, respectively. Panels (d–e) show the corresponding pointwise prediction errors. In both ROMs, $r=10$ modes were used; for the ql-ROM, six clusters were considered. Dashed lines indicate the prediction horizon. 
    }
	\label{fig:KScompar}
\end{figure}

Figure \ref{fig:KScompar} presents a comparison between the ground truth (test dataset) in panel (a), the g-ROM in panel (b), and the ql-ROM in panel (c). 
Panels (d) and (e) show the absolute value of the local error. Given the integration time-step $\Delta t$, the prediction horizon, defined as
 $T_{ph} = N_{ph}\Delta t$, in which $N_{ph}$ satisfies
 \begin{equation}\label{eq:predhor}
 	\|\bm{u}(N_{ph} \Delta t ) - \bm{u}_r(N_{ph} \Delta t ) \|  < \tau\sqrt{\frac{1}{N_{ph}}  \sum_{i=0}^{N_{ph}}  \|\bm{u}(i\Delta t ) \|^2   },
 	\end{equation}
 with $\bm{u}_r$ being the ql-ROM solution and $\tau=0.5$ as in \cite{Vlachas2020,Oezalp2023}, is indicated by dashed lines, showing the time span over which the ROMs can accurately predict the system's behavior. The ql-ROM has improved accuracy in capturing the nonlinear dynamics of the KS system compared to the g-ROM by a factor $\approx3$.

\begin{figure}
	\centering
	\subfloat[\phantom{a}\hfill\phantom{a}]{\centering\includegraphics[trim= 0cm 0cm 0cm 0cm,clip,width=0.5\columnwidth]{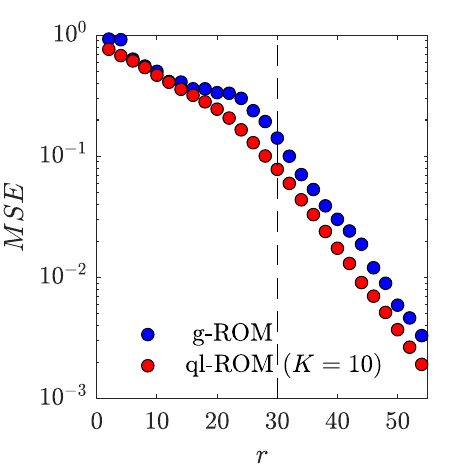}   }
	\subfloat[\phantom{a}\hfill\phantom{a}]{\centering\includegraphics[trim= 0cm 0cm 0cm 0cm,clip,width=0.5\columnwidth]{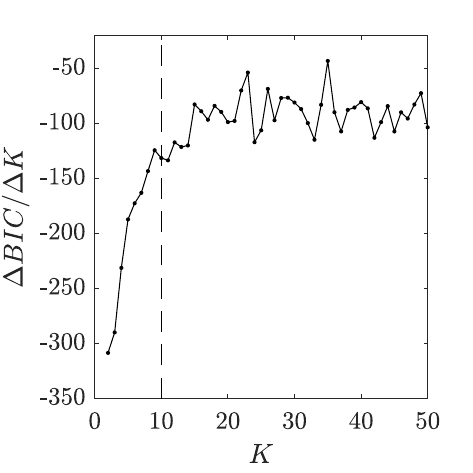}   }
	\caption{Parameters selection for the KS equations in the chaotic regime. Panel (a): reconstruction error of the test dataset for g-ROM (black) and ql-ROM (red). Panel (b): marginal variation of BIC score with the number of clusters $K$.}
	\label{fig:BICks2}
\end{figure}

\begin{figure}
	\centering
	\subfloat[\phantom{a}\hfill\phantom{a}]{\centering\includegraphics[trim= 0cm 0.4cm 0cm 1cm,clip,height=8.57cm]{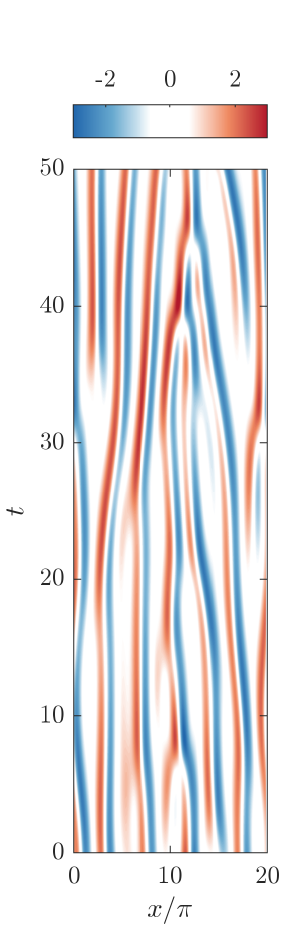}}
	\subfloat[\phantom{a}\hfill\phantom{a}]{\centering\includegraphics[trim= 1.08cm 0.4cm 0cm 1cm,clip,height=8.57cm]{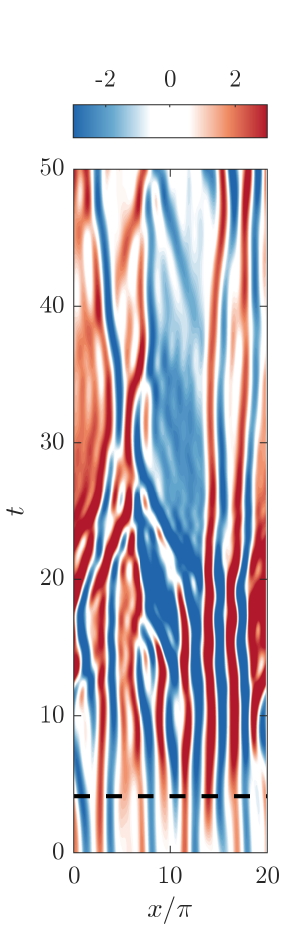}}
	\subfloat[\phantom{a}\hfill\phantom{a}]{\centering\includegraphics[trim= 1.08cm 0.4cm 0cm 1cm,clip,height=8.57cm]{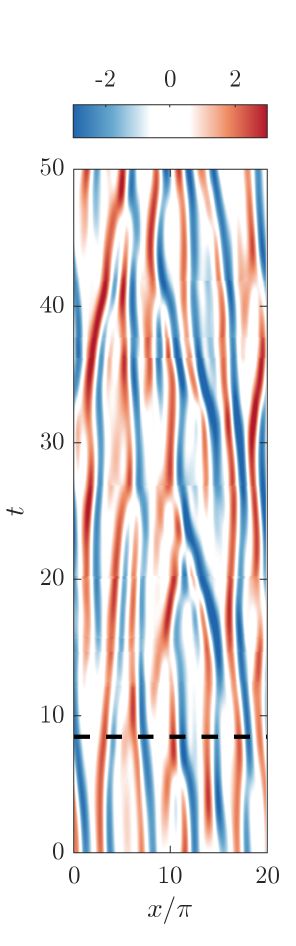}}
	\subfloat[\phantom{a}\hfill\phantom{a}]{\centering\includegraphics[trim= 1.08cm 0.4cm 0cm 1cm,clip,height=8.57cm]{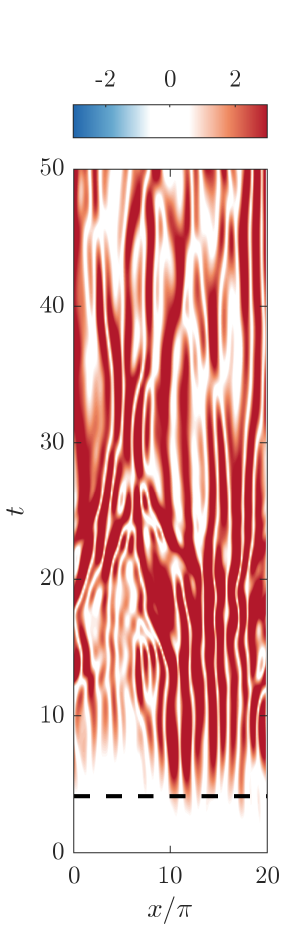}}
	\subfloat[\phantom{a}\hfill\phantom{a}]{\centering\includegraphics[trim= 1.08cm 0.4cm 0cm 1cm,clip,height=8.57cm]{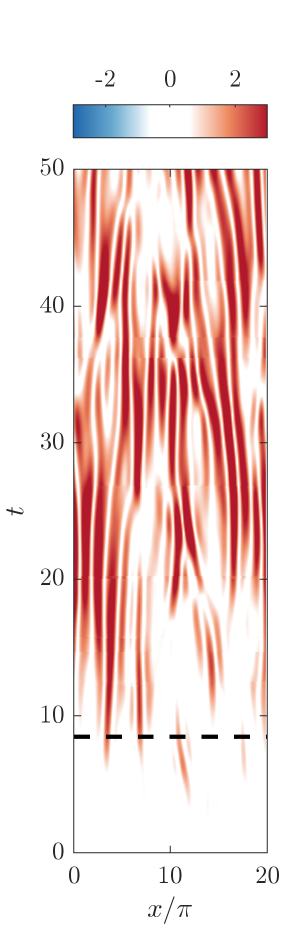}}\\

	\subfloat[\phantom{a}\hfill\phantom{a}]{\centering\includegraphics[trim= 0cm 0cm 0cm 0.cm,clip,width=0.3333\columnwidth]{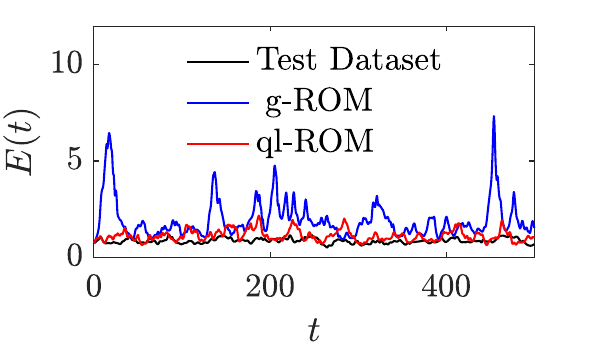}}
	\subfloat[\phantom{a}\hfill\phantom{a}]{\centering\includegraphics[trim= 0cm 0cm 0cm 0.cm,clip,width=0.3333\columnwidth]{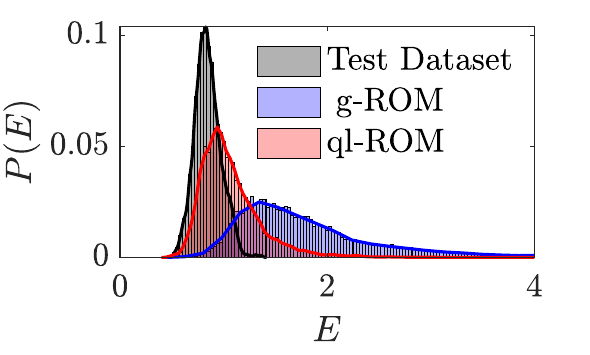}}
	\subfloat[\phantom{a}\hfill\phantom{a}]{\centering\includegraphics[trim= 0cm 0cm 0cm 0.cm,clip,width=0.3333\columnwidth]{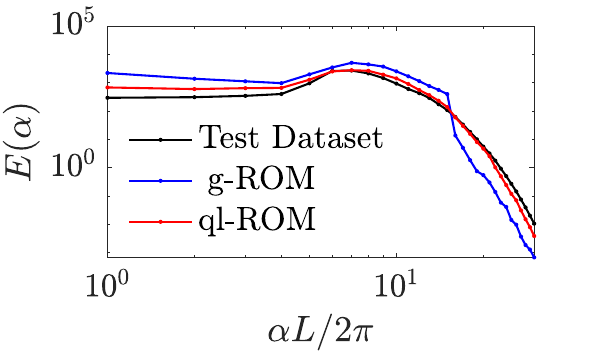}}

	\caption{Kuramoto–Sivashinsky dynamics in the chaotic regime. Panels (a–c): Ground truth, g-ROM prediction, and ql-ROM prediction. Panels (d–e): Corresponding errors. In both ROMs, $r=30$ modes are employed; for the ql-ROM, ten clusters are used. Dashed lines indicate the prediction horizon \eqref{eq:predhor}. Panel (f): Time evolution of kinetic energy $E(t)=\frac{1}{2L}\int_{\Omega}  \| u (x,t)\|^2 dx$. Panel (g): Long-term probability distribution of $E$.  The distributions (solid lines) are estimated using kernel density estimation (KDE) \cite{Scott1992}. Panel (h): Comparison between spatial energy spectra $E(\alpha)=\frac{1}{T}\int_{0}^T  \| \hat u (\alpha,t)\|^2 dt$, with $\hat u (\alpha,t)$ being the $\alpha^{th}$ spatial Fourier component of $u$.}
	\label{fig:KS2}
\end{figure}

A similar analysis has been carried out in the chaotic regime associated with $L = 20\pi$ and $\nu = 1$. The first step of the analysis is to choose $r$ and $K$. Figure \ref{fig:BICks2} shows the reconstruction error in panel (a) and the BIC's marginal variation $\Delta BIC/\Delta K$ in panel (b). $30$ modes provides an error that is lower than $10\%$. $K=10$ is the number of clusters at the elbow in panel (b).
A comparison between the predictions of the ROMs is shown in Panels (a-e) of Figure \ref{fig:KS2}. As in the bursting case, the ql-ROM  has greater accuracy in terms of prediction horizon also in chaotic regime by increasing it by a factor $\approx 2$.

Other quantities to analyse are the long term statistics, which can be represented by the kinetic energy $E(t)$ and the energy spectrum $E(\alpha)$ (defined in the caption of Figure \ref{fig:KS2}), $\alpha$ being a spatial Fourier wavenumber. Panel (f) of Figure \ref{fig:KS2} shows $E(t)$. Panel (g) shows the long-term probability distributions of $E$ for the various models, with the ql-ROM's distribution closely matching that of the test dataset. In panel (h), a comparison of $E(\alpha)$ across the models is depicted. The g-ROM spectrum presents aliasing at high wavenumbers while the ql-ROM closely aligns with the ground truth across all spatial scales. In conclusion, ql-ROMs prove to be a more robust and effective choice than g-ROMs with the same degrees of freedom. Users can expect improved numerical stability, a longer prediction horizon, and more accurate long-term statistics.

\subsection{Kolmogorov flow}\label{sec:resKOL}
The ql-ROM (Section \ref{sec:methods}) is applied to Kolmogorov flow which is higher-dimensional and multiscale. Two nonlinear regimes are analysed, the quasiperiodic regime, with $Re=20$, and the chaotic regime, with $Re=42$.

In the quasiperiodic case, the training dataset consists of an ensemble of $M=10^5$ snapshots of velocity components in the Fourier space with a time step  $\Delta t=0.1$, corresponding to a temporal simulation window $T_\mathrm{training} = 10000$.  The test dataset consists of $M_\mathrm{test}= 10^5$ snapshots with the same time step starting from the last snapshot of the training dataset. For this case the number of POD modes is equal to $r=100$ to ensure that the reconstruction error of the test dataset is lower than $0.1\%$. The number of clusters $K=10$ is chosen as detailed in Appendix \ref{sec:KolK}.

In Figure \ref{fig:Ko20}, a comparison between g-ROM and ql-ROMs is presented. 
	\begin{figure}
	\centering
\begin{minipage}{0.49\columnwidth}
\subfloat[\phantom{a}\hfill\phantom{a}]{\centering	\includegraphics[trim= 0cm 0cm 0cm 0.cm,clip,width=1\columnwidth]{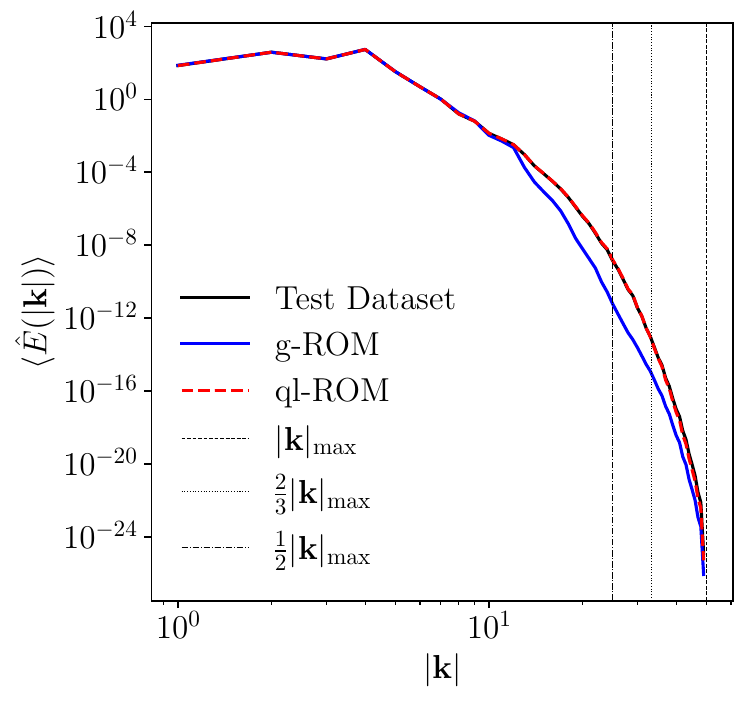} }  
\end{minipage}
\begin{minipage}{0.49\columnwidth}
\subfloat[\phantom{a}\hfill\phantom{a}]{\centering	\includegraphics[trim= 0cm 1.1cm 0cm 0cm,clip,width=1\columnwidth]{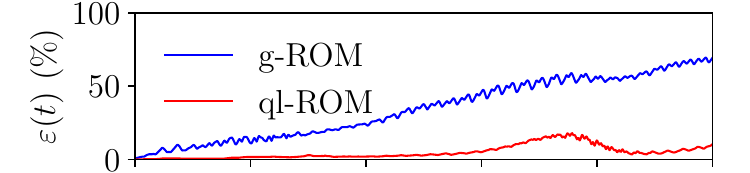} }\\
\subfloat[\phantom{a}\hfill\phantom{a}]{\centering	\includegraphics[trim= 0cm 1.1cm 0cm 0.cm,clip,width=1\columnwidth]{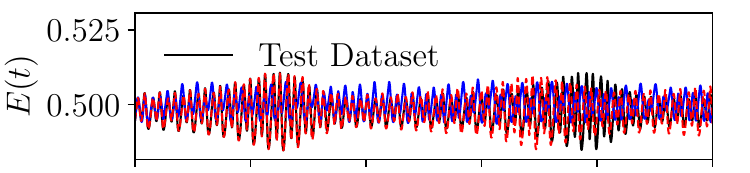}  } \\
\subfloat[\phantom{a}\hfill\phantom{a}]{\centering	
\includegraphics[trim= 0cm 0cm 0cm 0.cm,clip,width=1\columnwidth]{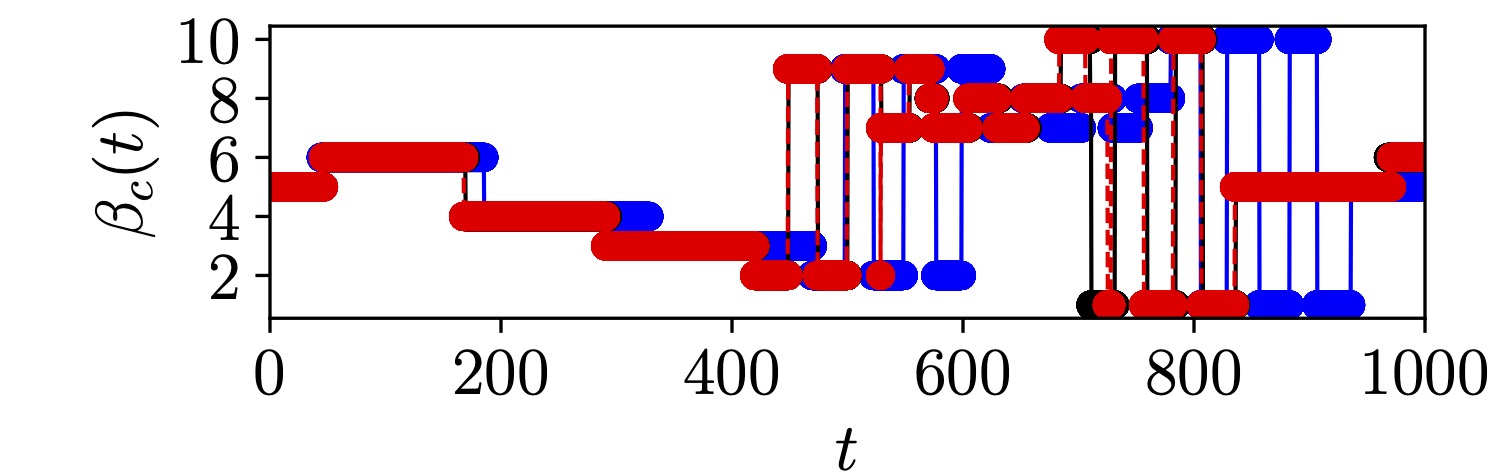} 
}  	\\
	\end{minipage}

\subfloat[\phantom{a}\hfill\phantom{a}]{\centering 	\includegraphics[trim= 0cm 0cm 0cm 0.cm,clip,width=1\columnwidth]{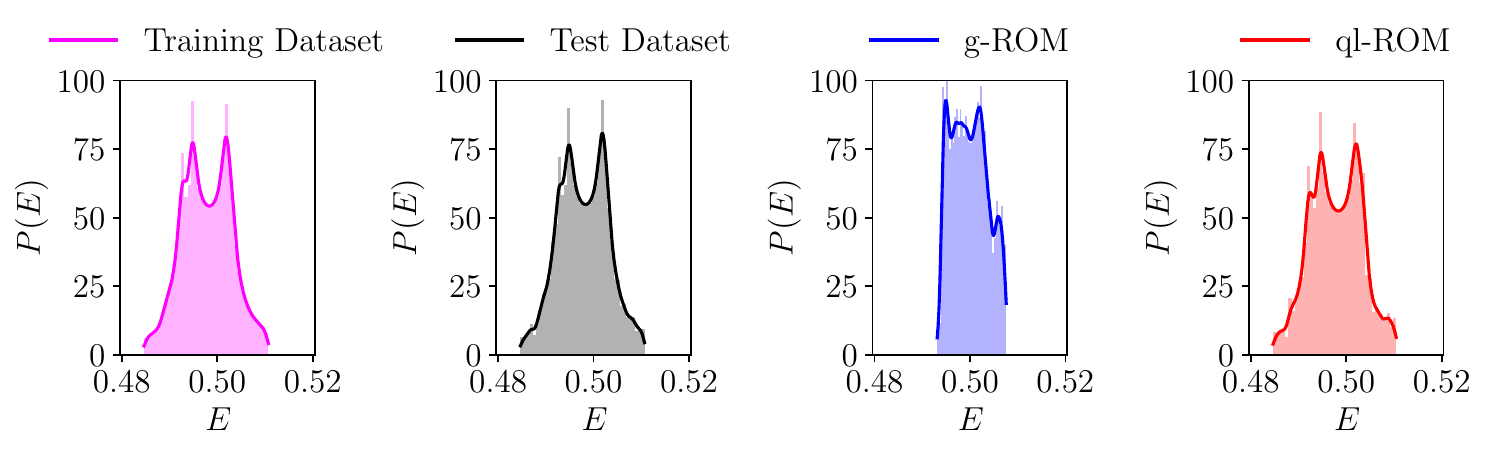}  
    }
	\caption{Kolmogorov flow. Comparison of g-ROM and ql-ROM predictions for the test dataset for quasiperiodic regime ($Re=20$). 
		Panel (a): Spatial energy spectrum $\langle E(|\mathbf{k}|) \rangle$. 
		Panel (b): Prediction error $\varepsilon(t)$. 
		Panel (c): Kinetic energy $E(t)$. 
		Panel (d): Cluster affiliation function, $\beta_c(t) = \beta(\mathbf{u}_r(t))$. Panel (e):  probability distribution functions (PDF) of the kinetic energy for the training dataset (magenta), test dataset (black), g-ROM (blue), and ql-ROM (red). The distributions (solid lines) are estimated using kernel density estimation (KDE) \cite{Scott1992}.}
	\label{fig:Ko20}
\end{figure}
The energy spectrum $\langle \hat{E}(|\mathbf{k}|) \rangle$ of the test dataset (panel (a)) has the characteristics of a direct energy cascade observed in turbulent flows, a multiscale phenomenon in which energy content decays with increasing wavenumber. The g-ROM spectrum deviates from the test case spectrum, particularly at high wavenumbers, failing to resolve higher wave numbers, which shows the limitations of the g-ROM. On the other hand, the ql-ROM, despite the same number of degrees of freedom, captures the spectrum that aligns closely with the ground truth. Panel (b) shows the prediction error for the test dataset
\begin{equation}
	\varepsilon (t) = \frac{\| \mathbf{u}(t) - \mathbf{u}_r(t) \|}{\| \mathbf{u}(t) \|}.
\end{equation}
The ql-ROM outperforms the g-ROM in terms of prediction error. Panel (c) shows the kinetic energy, $E(t)$, for the prediction dataset. The ql-ROM captures the multi-frequency behavior of the dataset.
Panel (d) shows the cluster affiliation function over time. The ql-ROM captures the transitions between clusters, accurately reproducing the temporal evolution of the cluster affiliation function. This shows that the ql-ROM explores the same regions of the phase space as the test dataset over time.
The  probability distribution functions (PDF) of the kinetic energy for the various datasets is shown in Panel (e) in Figure \ref{fig:Ko20}.
Bandwidths for the kernel density estimation (KDE) are determined using Scott’s rule \cite{Scott1992}. The training and test datasets exhibit the same probability distribution, and the ql-ROM closely reproduces this distribution.

We focus now on the chaotic configuration with $Re = 42$. For this setup, the training dataset consists of  $M = 8 \times 10^5$ snapshots, sampled at a time interval of  $dt = 0.05$. The test dataset contains  $M_{\mathrm{test}} = 2 \times 10^5$  snapshots with the same sampling interval. This configuration (Figure \ref{fig:komolgorovBASE}) evolves chaotically both in space and time. To ensure a low reconstruction error for the test dataset, the number of modes was increased to $r = 400$. The number of clusters is $K = 20$  according to the BIC score (Appendix \ref{sec:KolK}).

\begin{figure}
	\centering
	\setlength{\tabcolsep}{3pt}
	\renewcommand{\arraystretch}{0.1} 
	\begin{tabular}{ccccc}
		\vspace{8pt}
		& \multicolumn{2}{c}{g-ROM} &  \multicolumn{2}{c}{ ql-ROM}  \\
		$t=0$	& &  error & &  error\\
               \begin{overpic}[trim= 0cm 0cm 1.8cm 0cm,clip,height=2.32cm]{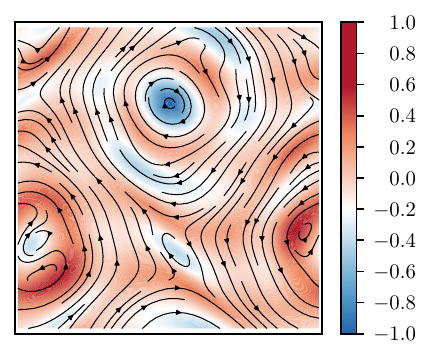}
      \put(-4,32){\scriptsize{$y$}} 
    \end{overpic} &
		\includegraphics[trim= 0cm 0cm 1.8cm 0cm,clip,height=2.32cm]{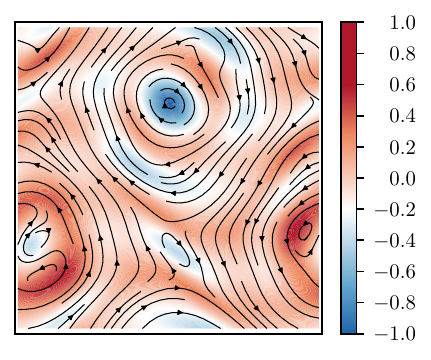}&
		\includegraphics[trim= 0cm 0cm 1.8cm 0cm,clip,height=2.32cm]{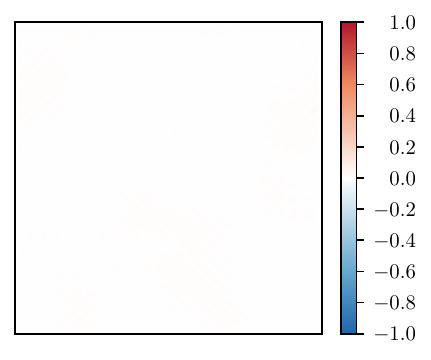}&
		\includegraphics[trim= 0cm 0cm 1.8cm 0cm,clip,height=2.32cm]{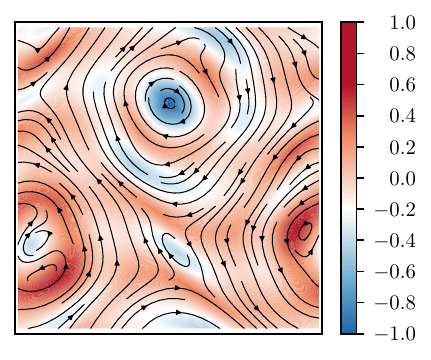}&
		\includegraphics[trim= 0cm 0cm 0.2cm 0cm,clip,height=2.32cm]{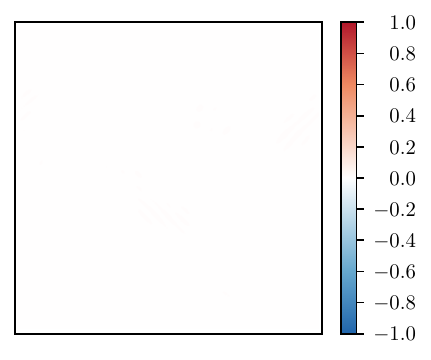}\\
		$t=20$& & \\

            \begin{overpic}[trim= 0cm 0cm 1.8cm 0cm,clip,height=2.32cm]{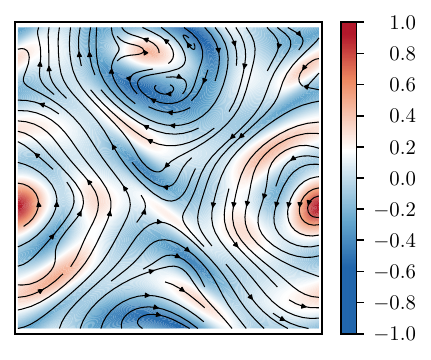}
      \put(-4,32){\phantom{\scriptsize{$y$}}} 
    \end{overpic}
        &
		\includegraphics[trim= 0cm 0cm 1.8cm 0cm,clip,height=2.32cm]{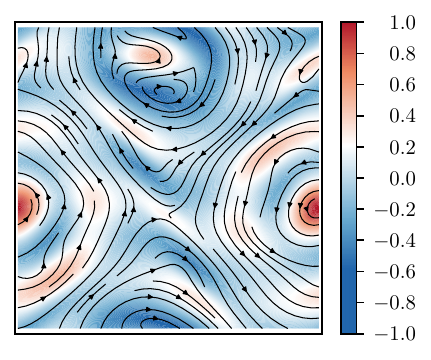}&
		\includegraphics[trim= 0cm 0cm 1.8cm 0cm,clip,height=2.32cm]{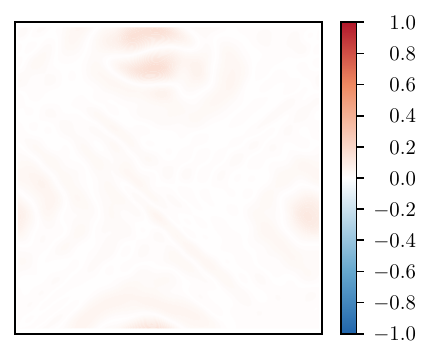}&
		\includegraphics[trim= 0cm 0cm 1.8cm 0cm,clip,height=2.32cm]{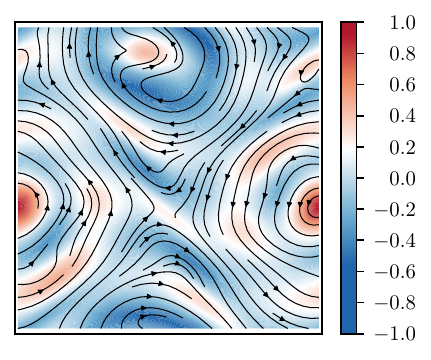}&
		\includegraphics[trim= 0cm 0cm 0.2cm 0cm,clip,height=2.32cm]{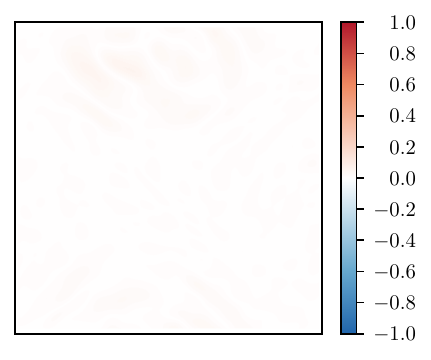}\\
		$t=40$& & \\
          \begin{overpic}[trim= 0cm 0cm 1.8cm 0cm,clip,height=2.32cm]{fig/nk35_nf4_RE42_Trans10000_Ttrain40000_results2snap800.pdf}
      \put(-4,32){\phantom{\scriptsize{$y$}}} 
    \end{overpic}
        
        &
		\includegraphics[trim= 0cm 0cm 1.8cm 0cm,clip,height=2.32cm]{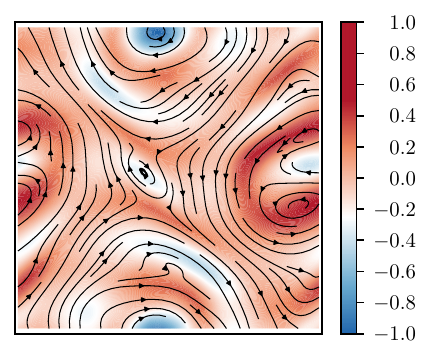}&
		\includegraphics[trim= 0cm 0cm 1.8cm 0cm,clip,height=2.32cm]{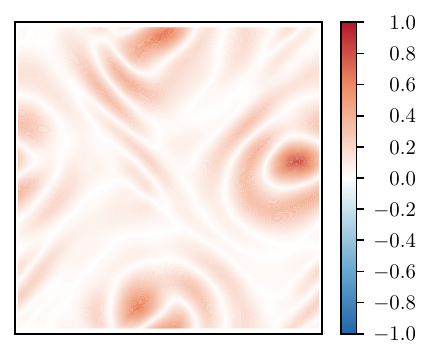}&
		\includegraphics[trim= 0cm 0cm 1.8cm 0cm,clip,height=2.32cm]{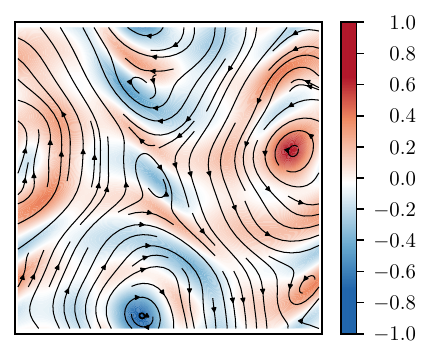}&
		\includegraphics[trim= 0cm 0cm 0.2cm 0cm,clip,height=2.32cm]{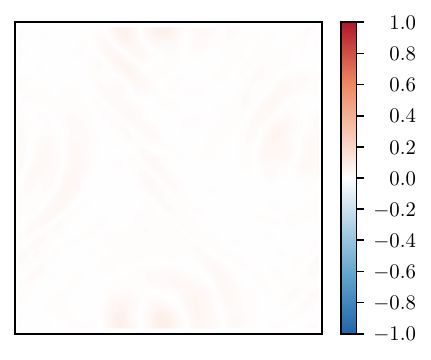}\\
		$t=60$& & \\

              \begin{overpic}[trim= 0cm 0cm 1.8cm 0cm,clip,height=2.32cm]{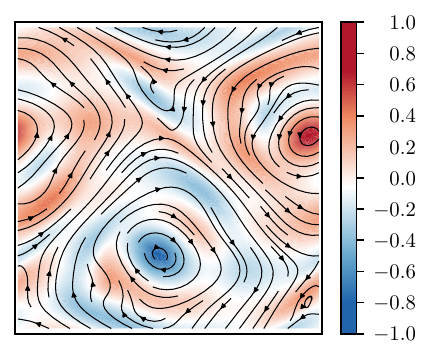}
      \put(-4,32){\phantom{\scriptsize{$y$}}}
       \put(30,-2){\scriptsize{$x$}}
    \end{overpic} 
        
        &
		\includegraphics[trim= 0cm 0cm 1.8cm 0cm,clip,height=2.32cm]{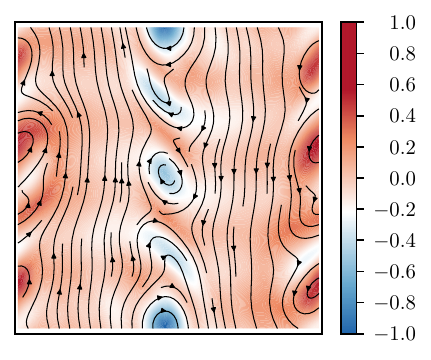}&
		\includegraphics[trim= 0cm 0cm 1.8cm 0cm,clip,height=2.32cm]{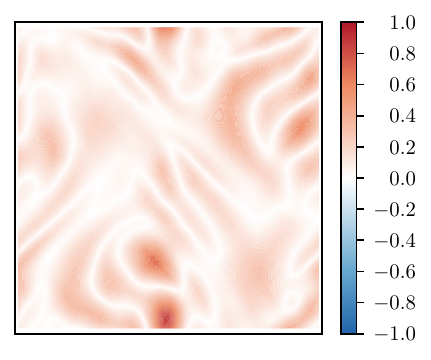}&
		\includegraphics[trim= 0cm 0cm 1.8cm 0cm,clip,height=2.32cm]{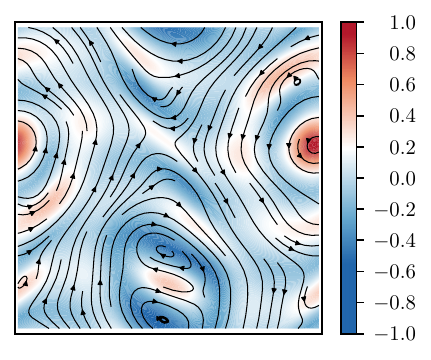}&
		\includegraphics[trim= 0cm 0cm 0.2cm 0cm,clip,height=2.32cm]{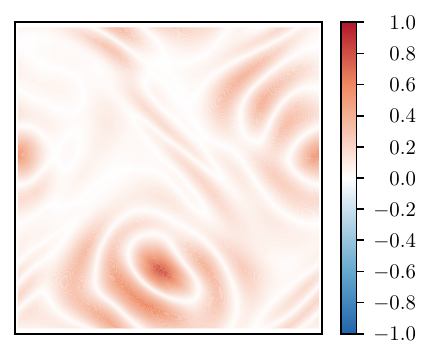}\\
		
	\end{tabular}

\caption{Kolmogorov flow at $Re=42$. Vorticity fields at four different time instances for the test dataset (first column). Predictions from the g-ROM (second column) and the ql-ROM (fourth column). The absolute value of the local error is shown in the third and fifth columns. In all the panels $0 < x < 2\pi$ and $0 < y < 2\pi$. All variables in each row are normalized with respect to the maximum vorticity value observed across the three snapshots shown.}
	\label{fig:Re42snaps}
\end{figure}

In Figure \ref{fig:Re42snaps}, the vorticity at four different time instances for the test dataset is shown with the corresponding predictions from g-ROM and ql-ROMs. 
At the initial prediction time, $t = 0$, before the ROM is deployed, the spatial distribution of the error is nearly negligible, as it arises only from the low reconstruction error \eqref{eq:residuo} associated with the choice of the number of modes used in constructing the ROM.
The comparison of the third and fifth columns, which display the local error over the time, shows that the ql-ROM is more accurate in predicting the flow dynamics.

\begin{figure}
	\centering

    \begin{minipage}{0.60\columnwidth}
        \subfloat[\phantom{a}\hfill\phantom{a}]{\centering 		\includegraphics[trim= 0cm 1.1cm 0cm 0.cm,clip,width=0.49\columnwidth]{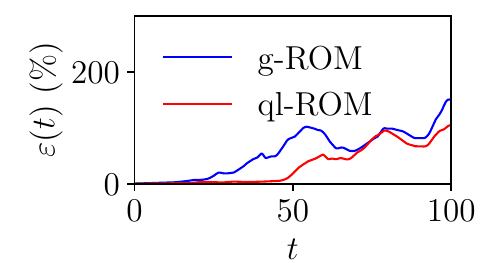}   
	\includegraphics[trim= 0cm 1.1cm 0cm 0.cm,clip,width=0.49\columnwidth]{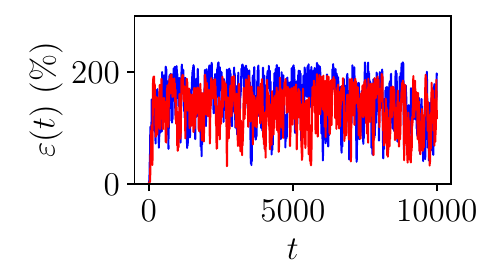}   
}	\\
\subfloat[\phantom{a}\hfill\phantom{a}]{\centering
	\includegraphics[trim= 0cm 0cm 0cm 0.cm,clip,width=0.49\columnwidth]{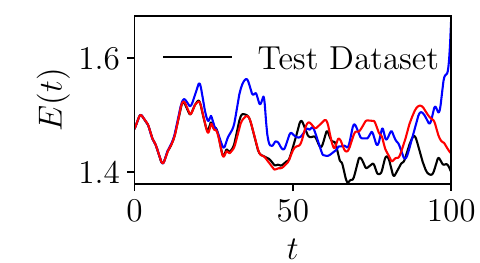}   
	\includegraphics[trim= 0.cm 0cm 0cm 0.cm,clip,width=0.49\columnwidth]{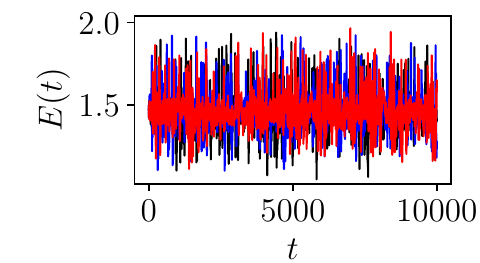}  
}
    \end{minipage}
   \begin{minipage}{0.39\columnwidth}
\subfloat[\phantom{a}\hfill\phantom{a}]{\centering 	
	\includegraphics[trim= 0cm 0cm 0cm 0.cm,clip,width=1\columnwidth]{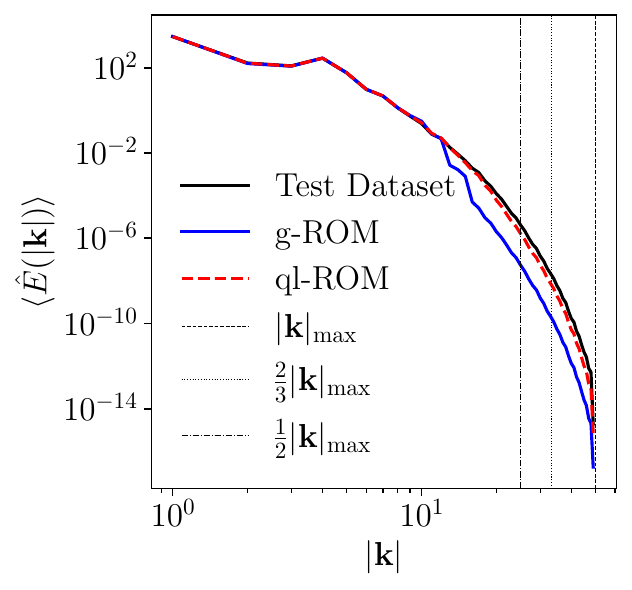}  
}
    \end{minipage}

\subfloat[\phantom{a}\hfill\phantom{a}]{\centering 	
\includegraphics[trim= 0cm 0cm 0cm 0.cm,clip,width=1\columnwidth]{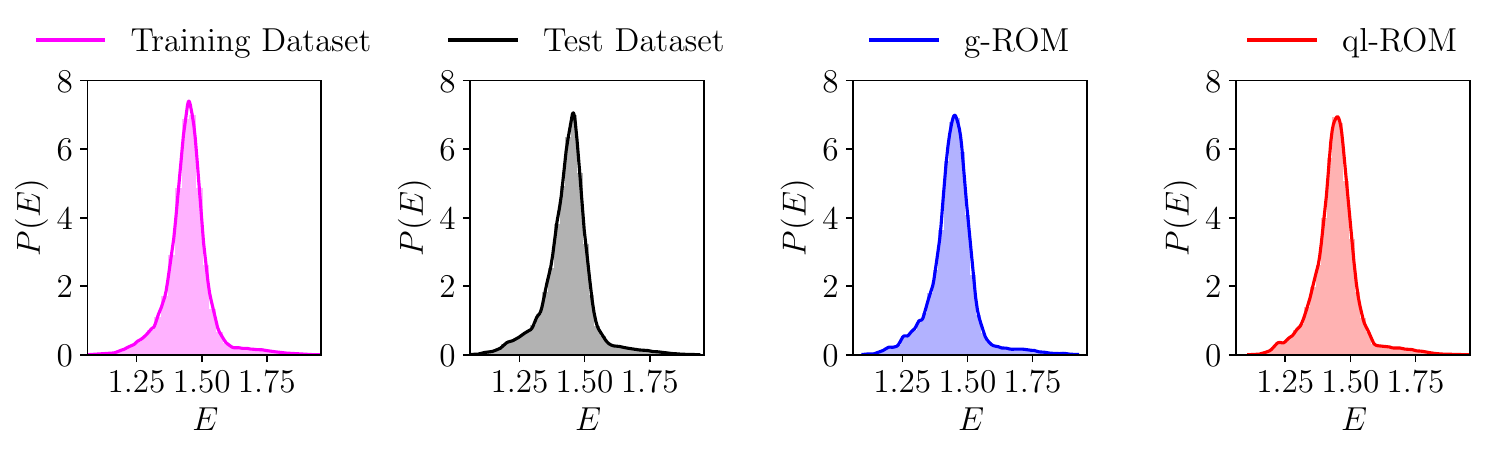}   

}

\caption{Kolmogorov flow. Comparison of g-ROM and ql-ROM predictions for the test dataset for chaotic regime ($Re=42$). Panel (a): Prediction error $\varepsilon(t)$ . The left panel shows a zoomed-in view for $0 < t < 100$, highlighting the lower prediction error of the ql-ROM. The right panel extends the time range to $t = 10,000$, showing that both ROMs remain stable over time. Panel (b): Comparison of kinetic energy $E(t)$ predictions among models for $Re = 42$. The left panel shows a zoomed-in view ($0 < t < 100$) highlighting the short-term accuracy of the quantized local approach. The right panel displays the long-term evolution of $E$. 
Panel (c): Spatial energy spectrum, $\langle E(|\mathbf{k}|) \rangle$, comparison between g-ROM and ql-ROM predictions for the test dataset. Panel (d): Estimated PDFs of kinetic energy for the training dataset (magenta), test dataset (black), g-ROM (blue), and ql-ROM (red).}
	\label{fig:recoRe42}
\end{figure}

The reconstruction error $ \varepsilon(t)$ is shown in Panel (a) of Figure \ref{fig:recoRe42}.
Both the g-ROM and the ql-ROMs remain stable over time, as indicated by the error in both cases, but the ql-ROM has a lower prediction error.
Panel (b) of Figure \ref{fig:recoRe42} shows a comparison of kinetic energy $E(t)$ among the ROMs. Both g-ROM and ql-ROMs capture the intermittent and random bursts in kinetic energy but ql-ROM predicts accurately the short time behavior of $E(t)$.
Panel (c) of Figure \ref{fig:recoRe42} presents the mean spatial energy spectrum $ \langle \hat{E}(|\mathbf{k}|) \rangle $, for the different datasets. At high wavenumbers ($ |\mathbf{k}| > 12 $), the g-ROM has an energy content approximately two orders of magnitude lower than the original dataset. On the other hand, the ql-ROM has a spatial spectrum that closely matches the ground truth.
The estimated PDFs of the kinetic energy are shown in Panel (d) of Figure~\ref{fig:recoRe42}. Both models effectively reproduce the statistical characteristics of the flow, including the probability tails associated with the rare bursts of $E$. 

In conclusion, the ql-ROM outperforms the g-ROM in both the quasiperiodic and chaotic  regimes of Kolmogorov flow. It achieves lower prediction error, better reconstruction of kinetic energy, and more accurate spatial energy spectra. The ql-ROM also captures the statistical properties of the flow more faithfully than the g-ROM. These results demonstrate the improved accuracy and robustness of the ql-ROM in modeling complex, multiscale dynamics.

\section{Conclusions}\label{sec:concl}

In chaotic dynamical systems, the geometry of the attractor is often intricate and heterogeneous. This complexity poses a challenge for global reduced-order modeling (g-ROM), i.e., a single model may fail to capture the localized dynamics across the manifold, which may result in  loss of accuracy and numerical stability. To address this limitation, we  introduce a divide-and-conquer framework in time: quantized local reduced-order modeling (ql-ROM). 
The proposed methodology consists of three main steps. First, the solution manifold (data) is quantized into different regions with clustering techniques, which generates an approximate cartography. Second, local ROMs are constructed around the centroids of each cluster. Third, the most accurate local model is adaptively selected based on the closest cluster.
We employ the K-means algorithm for phase-space quantization and Galerkin projections of the governing equations for constructing the local ROM. The number of clusters and the number of retained modes (degrees of freedom) are selected with the Bayesian information criterion (BIC),  which provides a principled trade-off between model complexity and fidelity. The methodology is  intrusive, deterministic, and physically interpretable because it originates from the governing equations. 
The ql-ROMs are constructed and tested on two  nonlinear partial differential equations: the Kuramoto–Sivashinsky equation and the Navier-Stokes equation, both of which have multiscale and spatiotemporal chaotic dynamics. The ql-ROMs significantly and consistently outperforms g-ROMs with the same number of degrees of freedom. This improvement is assessed  with different metrics, from short-term prediction,  through long-term statistics, to spectral content, and the model's numerical stability. 
The computational overhead of ql-ROMs is minimal. ql-ROMs open opportunities for nonlinear model reduction in time, data assimilation, among others. 

\section*{Acknowledgments}
We acknowledge the support from the grant EU-PNRR YoungResearcher TWIN ERC-PI\_0000005.  

		



	
	\appendix

\section{Nomenclature}
\label{sec:variables}

For clarity and reference, Table~\ref{tab:variables} provides a summary of the main symbols and variables used throughout the manuscript, along with their definition.

\begin{table}
\centering
\def~{\hphantom{0}}
  \begin{tabular}{ll}
        Variables   &   Description\\[0.5pt]
        $\bm{u}_m$    &   Time-resolved snapshots\\[0.5pt]
          $\bm{q}_m$    &   Time-resolved snapshots in wave-numbers domain\\[0.5pt]
        $m$    &   Snapshots index\\[0.5pt]
        $M$   &   Number of training dataset snapshots\\[0.5pt]
         $M_{\mathrm{test}}$   &   Number of test dataset snapshots\\[0.5pt]
        $t$   &   Time\\[0.5pt]
        $\Delta t$   &  Time step\\[0.5pt]
        $\bm{\mu}$ &  Mean field\\[0.5pt]
        \multicolumn{2}{l}{ -----------------  Clustering ----------------- }\\[0.5pt]  
        $K$   &   Number of clusters\\[0.5pt]
        $\mathcal{C}_{k}$   &   Clusters \\[0.5pt]
        $\chi_{i}^{m}$   &   Characteristic function of the state space clustering\\[0.5pt]
            $n_{k}$   &   Number of snapshots in cluster $\mathcal{C}_{k}$\\[0.5pt]
         $\beta_v(\bm{u})$ &   Cluster-affiliation function for state vector\\[0.5pt]
         $\beta_c(t)$ &   Cluster-affiliation function in time\\[0.5pt]
        $\beta(m)$ &   Cluster-affiliation function for time index\\[0.5pt]
         $d_{m,n}^2$   &   Squared Euclidean distance between two points $m,n$\\[0.5pt]
        $\boldsymbol{c}_{k}$   &   Centroids of clusters \\[0.5pt]
        $J$&  Intra-cluster variance\\[0.5pt]

        \multicolumn{2}{l}{ ----------------- Quantized local Galerkin POD ROM ----------------- }\\[0.5pt] 
        $\bm{u}'_m$    &   Fluctuation with respect to the nearest centroid\\[0.5pt]
      $\mathbf{Q'}_k$&   Matrix of the snapshots belonging to cluster $k$\\[0.5pt] 
  $\mathbf{U}_k$&   Matrix of the spatial modes within the cluster $k$\\[0.5pt] 
 $\mathbf{V}_k$&   Matrix of the temporal modes within the cluster $k$\\[0.5pt] 
  $\mathbf{\Sigma}_k$&   Matrix of the singular values within the cluster $k$\\[0.5pt]  
     $r_k$&   Number of retained modes for the cluster $k$\\[0.5pt] 
  $\bm{\varphi}_i^k$&  Spatial mode within the cluster $k$\\[0.5pt]      
      $\bm{r}$&   Reconstruction error\\[0.5pt] 
          BIC&   Bayesian information criterion score\\[0.5pt]  
   $n_p$&   Number of parameters in the model\\[0.5pt]

  \end{tabular}
\caption{Table of variables.}
\label{tab:variables}
\end{table}

\section{Numerical Treatment of the analysed Testcases}
\label{sec:numerics}

In practice, the governing equation \eqref{eq:dyn} is often discretized numerically in the Fourier spectral domain \cite{Canuto2007}. This results in the equivalent formulation
\begin{equation}
	\frac{\partial {\bm{q}}}{\partial t} + \hat{\mathcal{N}}({\bm{q}},t) = 0, \quad {\bm{q}} \in \mathbb{C}^{N},
\end{equation}
where ${\bm{q}}(\mathbf{k}, t)$ and $\hat{\mathcal{N}}$ represent the spectral counterparts of $\bm{u}( t)$ and the nonlinear operator $\mathcal{N}$, respectively, and $\mathbf{k}$ is the spatial wavenumber vector. All numerical simulations and data analyses in this work are conducted in the spectral domain.

\subsection{Kuramoto–Sivashinsky Equation}
The KS equation \eqref{eq:ks} is rewritten as:
\begin{equation*}
	u_t = Lu + N(u),
\end{equation*}
where the linear term is $Lu = -u_{xx} - \nu u_{xxxx}$, and the nonlinear term is $N(u) = -u u_x$. The system is integrated in time using a fourth-order exponential time differencing Runge–Kutta method (ETDRK4) combined with a spectral discretization \cite{Cox2002}. The spatial domain is discretized using $n_x = 128$ equispaced points, corresponding to the same number of Fourier modes. The time step is set to $\Delta t = 0.05$, which satisfies the Courant–Friedrichs–Lewy (CFL) condition \cite{Lele1992}. The initial condition is set to $u(x,0) = \cos(x)$. A transient of $T_{tr} = 1.5 \times 10^3$ time units is discarded in all analyses to ensure statistical stationarity.

\subsection{Kolmogorov flow }
The Kolmogorov flow is solved using a differentiable pseudospectral method. The spatial discretization is performed via Fourier transforms, such that $\mathbf{q} = \mathcal{F} \circ \mathbf{u}$, where $\mathcal{F}$ denotes the Fourier transform and $\mathbf{q} \in \hat{\Omega}_k \subset \mathbb{C}^n$ is the spectral representation of the velocity field. In the Fourier domain, the incompressibility constraint is automatically satisfied \cite{Canuto2007, Canuto1988}. The evolution equation becomes:
\begin{equation*}
	\left( \frac{d}{dt} + \nu |\mathbf{k}|^2 \right) \mathbf{q}_k - \hat{\mathbf{f}}_k + \mathbf{k} \frac{\mathbf{k} \cdot \hat{\mathbf{f}}_k}{|\mathbf{k}|^2} - \hat{\mathbf{g}}_k = 0,
\end{equation*}
where $\hat{\mathbf{f}}_k = -(\mathcal{F} \circ (\mathbf{u} \cdot \nabla \mathbf{u}))_k$ represents the nonlinear convective terms. Nonlinear terms are computed pseudospectrally, and the 2/3 dealiasing rule is used to prevent spectral aliasing errors \cite{Canuto1988}.
Time integration is performed using an explicit forward Euler scheme with a simulation time step $\Delta t_s$ chosen to satisfy the CFL condition. For $Re = 20$, $\Delta t_s = 0.01$, while for $Re = 42$, $\Delta t_s = 0.005$. Initial conditions are generated using random fields scaled by wavenumber to preserve multiscale spatial structure \cite{Ruan1998}.
To ensure statistical stationarity, an initial transient of $T_{\mathrm{tr}} = 10,000$ time units is discarded in both cases. The number of snapshots used in the training and test datasets, along with the sampling time step (which may differ from the simulation time step), is reported for each case in Section~\ref{sec:results}.

\subsection{Complex-valued phase space quantization}
Since the state vectors are obtained from spectral discretizations, they are complex-valued. To enable clustering and ROM construction, which require real-valued input, the state vectors are mapped to an equivalent real-valued representation
\begin{equation}
	\bm{\xi}_m = 
	\begin{bmatrix}
	\Re(\bm{q}_m) \\
	\Im(\bm{q}_m)
	\end{bmatrix},
\end{equation}
where $\Re(\cdot)$ and $\Im(\cdot)$ denote the real and imaginary parts, respectively.

The Euclidean distance between two complex-valued snapshots $m$ and $n$ is then defined as
\begin{equation}
	d_{m,n}^2 = (\bm{\xi}_m - \bm{\xi}_n)^H(\bm{\xi}_m - \bm{\xi}_n),
\end{equation}
where $(\cdot)^H$ denotes the Hermitian transpose. This formulation allows standard clustering algorithms and reduced-order modeling techniques to be directly applied in the transformed real-valued space.

\section{Classical multidimensional scaling (MDS)}\label{sec:MDS}

	Multidimensional scaling (MDS) aims to represent high-dimensional data in a low-dimensional space and to preserve the pairwise distances between points. We use MDS to visualize the high-dimensional flow states on a two-dimensional map, enabling the identification of the system's regime.
    
The pairwise distances of two snapshots $d_{m,n}$, computed as \eqref{distFT}, are stored into the matrix $\mathbf{D}$. Then, a matrix $\mathbf{A} = -\frac{1}{2} \mathbf{C} \mathbf{D}^{2} \mathbf{C}$ is constructed with the squared proximity matrix $\mathbf{D}^{2}$ and $\mathbf{C} =\mathbf{I} - \frac{1}{M} \mathbf{1} \mathbf{1}^T$, where $\mathbf{I}$ is the identity matrix of size $M \times M$, and $\mathbf{1}$ is a column array of all ones of length $M$, $M$ being here the total number of snapshots. In the end, the pairwise distances $D_{m,n}$ can be represented in a feature space $\gamma = [\gamma_1, \gamma_2, \dots, \gamma_N]$, where the elements of $\gamma$ are ordered by their contribution to the distance measurement.

	Here, we chose a two-dimensional subspace that approximates $\tilde{\gamma} \approx [\gamma_1, \gamma_2]$ for visualization. We then determine $[\gamma_1, \gamma_2] = V \Lambda^{1/2}$, where $\Lambda$ and $V$ contain the first two eigenvalues and eigenvectors of $A$. The proximity map $[\gamma_1, \gamma_2]$ is the optimal plan that preserves as much as possible the distances in the original high-dimensional space.
    
\section{Phase space quantization of the Kolmogorov flow.}\label{sec:KolK}	
The construction of quantized local ROMs for the Kolmogorov flow requires the phase space to be partitioned into $K$ clusters. Selecting an appropriate value for $K$ is crucial to balance model complexity and representational accuracy. In this work, the number of clusters was determined using an elbow method applied to the Bayesian information criterion (BIC) score, as described in Section~\ref{sec:choicerK}.

\begin{figure}
	\centering
	\subfloat[\phantom{a}\hfill\phantom{a}]{%
		\includegraphics[trim=0cm 0cm 0cm 0cm, clip, width=0.48\columnwidth]{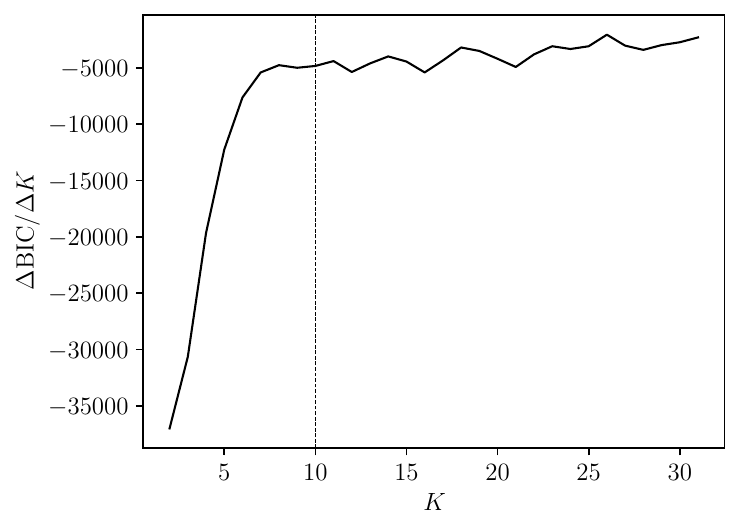}
	}
	\subfloat[\phantom{a}\hfill\phantom{a}]{%
		\includegraphics[trim=0cm 0cm 0cm 0cm, clip, width=0.48\columnwidth]{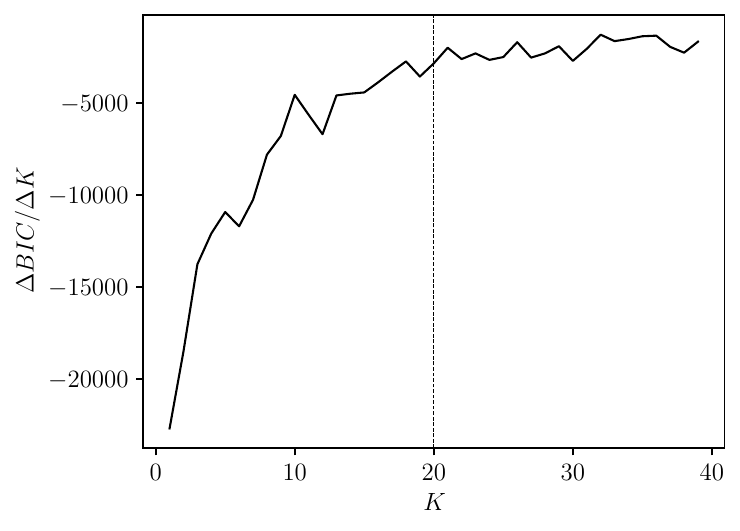}
	}
	\caption{Marginal variation of the BIC score used to determine the number of clusters $K$ for the Kolmogorov flow. Panel (a): quasi-periodic regime ($Re = 20$), where the elbow indicates $K = 10$. Panel (b): chaotic regime ($Re = 42$), where a more gradual decay in the BIC score suggests selecting $K = 20$.}
	\label{fig:BICkolm}
\end{figure}

Figure~\ref{fig:BICkolm} shows the marginal decrement of the BIC score as a function of the number of clusters $K$ for the two Kolmogorov flow regimes analysed. 
For the quasi-periodic regime at $Re = 20$, an elbow in the BIC curve suggests selecting $K = 10$ clusters. In contrast, the more complex and chaotic regime at $Re = 42$ exhibits a more gradual BIC decay, with the most suitable trade-off occurring at around $K = 20$.

	\bibliography{references}

\end{document}